\documentclass{aa}  

\usepackage{graphicx}
\usepackage{txfonts}
\usepackage{makecell}
\usepackage{multirow}
\usepackage{threeparttable}
\usepackage{xcolor}
\usepackage{orcidlink}
\usepackage{color}
\usepackage{float}
\usepackage[switch,mathlines]{lineno}

\begin{document} 
% \linenumbers

   \title{Doppler imaging combined with high-cadence photometry. I. \\ Revisiting the surface of a pre-main-sequence flare star \\ 
   PW Andromedae}

   \subtitle{}

   \author{S. Lee
          \inst{1}\orcidlink{0009-0007-7374-5133},
          E. Bahar\inst{2}\orcidlink{0000-0001-9119-2316},
          H. V. \c{S}enavc{\i}\inst{2}\orcidlink{0000-0002-8961-277X},
          E. I\c{s}{\i}k\inst{3}\orcidlink{0000-0001-6163-0653},
          K. Ikuta\inst{4}\orcidlink{0000-0002-5978-057X},
          K. Namekata\inst{5,6,7,8,9}\orcidlink{0000-0002-1297-9485},
          H. Nagata\inst{10}, \\
          K. Kawauchi \inst{11}\orcidlink{0000-0003-1205-5108},
          M. Omiya\inst{12,9}\orcidlink{0000-0002-5051-6027},
          H. Izumiura\inst{13}\orcidlink{0000-0002-8435-2569},
          A. Tajitsu\inst{13}\orcidlink{0000-0001-8813-9338},
          B. Sato\inst{14}\orcidlink{0000-0001-8033-5633},
          S. Honda\inst{10}\orcidlink{0000-0001-6653-8741},
          \and
          D. Nogami\inst{1}\orcidlink{0000-0001-9588-1872}
          }

   \institute{Department of Astronomy, Kyoto University, Kitashirakawa-Oiwake-cho, Sakyo-ku, Kyoto, 606-8502, Japan,
              \\
              \email{sanghee@kusastro.kyoto-u.ac.jp}
         \and
             Department of Astronomy and Space Sciences, Faculty of Science, Ankara University, Be\c{s}evler 06100, Ankara, T\"urkiye\\
             \email{enbahar@ankara.edu.tr}
         \and
             Max-Planck-Institut f\"ur Sonnensystemforschung, Justus-von-Liebig-Weg 3, D-37077 G\"ottingen, Germany
        \and  
             Department of Social Data Science, Hitotsubashi University, 2-1 Naka, Kunitachi, Tokyo 186-8601, Japan
        \and 
             The Hakubi Center for Advanced Research, Kyoto University, Yoshida-Honmachi, Sakyo, Kyoto 606-8501, Japan
        \and
             Department of Physics, Kyoto University, Kitashirakawa-Oiwake-cho, Sakyo, Kyoto 606-8502, Japan
        \and 
             NASA Goddard Space Flight Center, 8800 Greenbelt Road, Greenbelt, MD 20771, USA
        \and 
             Department of Physics, The Catholic University of America, 620 Michigan Ave NE, Washington, DC 20064, USA
        \and 
             National Astronomical Observatory of Japan, NINS, 2-21-1 Osawa, Mitaka, Tokyo 181-8588, Japan
        \and
            Nishi-Harima Astronomical Observatory, University of Hyogo, 407-2 Nishigaichi, Sayo-cho, Sayo, Hyogo 679–5313, Japan
        \and
            Department of Physical Sciences, Ritsumeikan University, 1-1-1 Noji-higashi, Kusatsu, Shiga 525-8577, Japan
        \and
             Astrobiology Center, NINS, 2-21-1 Osawa, Mitaka, Tokyo 181-8588, Japan
        \and
             Okayama Branch Office, Subaru Telescope, National Astronomical Observatory of Japan, NINS, 3037-5 Honjo, Kamogata, Asakuchi, Okayama 719-0232, Japan
        \and
             Department of Earth and Planetary Sciences, Institute of Science Tokyo, 2-12-1 Ookayama, Meguro-ku, Tokyo 152-8551, Japan
            }

   \date{}

% \abstract{}{}{}{}{} 
% 5 {} token are mandatory
 
  \abstract
   {Latitude distribution of stellar magnetic activity is not well constrained by observations, despite its importance for a better understanding of stellar dynamos and their effects on planetary environments.}
   {We aim to obtain an accurate reconstruction of the surface spot distribution on the young, rapidly rotating K2 star PW Andromedae by combining spectroscopic and photometric diagnostics. In particular, we seek to assess how the inclusion of continuous high-precision TESS photometry in parallel with high-resolution spectroscopy improves latitude recovery of starspots, especially at low latitudes and in the southern hemisphere, which are poorly constrained by Doppler imaging (DI) alone. We thereby explore the spatial origins of the observed white-light flares.}
   {We performed simultaneous Doppler imaging and light‐curve inversion (DI+LCI) using contemporaneous high-resolution GAOES‐RV spectra from the 3.8 m Seimei telescope (\textit{R} $\sim$ 65000) and high-precision TESS light curves. Surface reconstructions employ the {\tt SpotDIPy} code to model both line profiles and continuum brightness variations. We compare DI+LCI maps with DI‐only solutions, conduct artificial‐spot simulations to evaluate the effects of latitude, phase coverage, and signal‐to‐noise ratio on reconstruction reliability. We also investigate the spatial correlation between the DI+LCI reconstructed map and flares detected in the TESS data.}
   {The DI+LCI reconstruction reveals significant spot features at mid-to-low latitudes, equatorial regions, and even in the southern hemisphere. These are the regions where DI-only fails to provide reliable information. Meanwhile, the high-latitude spot features, which are already recovered by DI-only, remain present, though with a restructured distribution. The estimated spot coverage is approximately 9.9\% of the area of the stellar surface visible to the observer. Simulations show that DI+LCI provides more accurate reconstructions than DI-only, especially under conditions of incomplete phase coverage and low signal-to-noise, by better recovering both spot latitudes and filling factors. A comparison between the DI+LCI map and the TESS flare timings also suggests potential association between flare occurrence and reconstructed spot longitudes.}
   {Simultaneous DI and continuous photometry improves inversion accuracy of starspot distributions, also elucidating flare localization.}

   \keywords{Stars: activity - Stars: imaging - Stars: individual: PW Andromedae - starspots - Stars: flare}
 \titlerunning{Simultaneous DI and LCI of PW And}
 \authorrunning{S. Lee et al.}
   \maketitle
%
%-------------------------------------------------------------------

\section{Introduction}\label{sec:intro}

Stellar magnetic activity, which includes phenomena such as starspots and flares, is most pronounced in young and rapidly rotating stars. These stars exhibit intense magnetic fields and significant variability in their activity cycles, making them crucial targets for understanding the mechanisms driving stellar dynamos and their influence on stellar evolution \citep[for a review,][]{Brun2017}. Studying their magnetic activity provides a unique opportunity to investigate how patterns of magnetic activity on the Sun developed in its early evolutionary stage \citep[e.g.,][]{Kriskovics2019}. Among various indicators of stellar magnetic activity, starspots are particularly significant, as their distribution and evolution offer key constraints on the structure and dynamics of stellar magnetic fields \citep[for reviews,][]{Berdyugina2005, Strassmeier2009}.
Importantly, the presence of large, complex starspot groups is often associated with strong magnetic activity, which can result in the occurrence of intense flaring events \citep[for a review,][]{Kowalski2024}. Observations of young, active stars have revealed that their large spots often exhibit superflare events with energies exceeding those of the largest solar flares with the energy of $\sim 10^{32}$ erg by more than an order of magnitude \citep[e.g.,][]{Notsu2013,Notsu2015,Maehara2017}. These solar and stellar studies suggest that the size, complexity, and distribution of starspots are critical factors in determining flare activity. Furthermore, as in the Sun, the spatial distribution of starspots and the surrounding magnetic field structure can be closely related to the occurrence of flares \citep{Namekata2024}. 
Young and active Sun-like stars exhibit a wide range of magnetic behaviors, from periodic activity cycles to more irregular variations  \citep[e.g.,][]{Baliunas1995, Lee2023,Lee2024}. The irregular variations in magnetic activity, including fluctuations in brightness or S-index, are more common in younger and more active stars \citep{Hawley2014}. Their surface spot configurations may often differ significantly from the Sun's, with many showing prominent high-latitude spots or multiple active longitudes \citep{Berdyugina2005}. 
While investigating the correlation between flare frequency and rotational phase can provide valuable insights into the spatial distribution and migration of starspots, the interpretation of phase-dependence of flares in young, active stars can be complex.
Therefore, mapping the surface distribution of starspots is essential for understanding its association with flare activity and the overall evolution of stellar magnetic fields \citep{Ikuta2023}, in addition to contributing studies of how the spot distribution changes with stellar parameters. 
   
To study the surface distribution of starspots, two primary techniques based on spectroscopic and photometric data are widely employed: Doppler imaging (DI) and light-curve inversion (LCI). DI has proven to be a powerful technique for reconstructing the surface distribution of starspots on rapidly rotating stars by analyzing time-resolved line profile distortions modulated by stellar rotation \citep[see][]{Strassmeier1998}. DI is particularly sensitive to high-latitude features due to the way rotational broadening affects spectral lines. Photometric brightness modulations, on the other hand, provide information about starspot distribution by measuring brightness variations caused by in and out of starspots to the line of the sight \citep{Ikuta2020}. Light curve inversion of spotted stars constrains the properties of starspots (such as longitudinal distribution, size, and temporal evolution), which provides valuable insights into stellar dynamos, and is especially effective at revealing the presence of spots at low- to mid-latitudes, where the rotational modulation of the light curve is most pronounced \citep{Finociety2021,Waite2011}.  However, each method has its intrinsic limitations. A simple analysis of light curves does not fully resolve spatial information, owing to the degeneracy between latitude, size, and contrast of spots, making it particularly difficult to accurately determine spot latitudes. While the longitude can be determined with high precision, especially in short-cadence space-based photometry, the latitude remains much less constrained. Likewise, DI alone struggles to accurately recover low-latitude spots, as it is more sensitive to features at higher latitudes with a dependence on the axial inclination of the rotation axis \citep{Jeffers2002}. These limitations highlight the importance of combining both methods to obtain a more complete picture of the surface distribution of starspots.

PW Andromedae (PW And), a pre-main-sequence star of spectral type K2, with a rotation period of 1.76 days and an estimated axial inclination of 46$^{\circ}$ \citep{Strassmeier2006}, serves as an ideal target for the DI study. Since young active stars exhibit strong magnetic behavior, investigating the surface activity of pre-main-sequence stars like PW And can provide valuable insights into stellar magnetic activity during this crucial stage of stellar evolution. The magnetic activity of PW And has been studied through the analysis of photospheric and chromospheric activity variations, revealing presence of cool spots and flare activity \citep{Lopez-Santiago2003}. Several DI studies have been conducted to map the star’s surface, and their results are often discussed in terms of spot latitude. The low-, mid-, and high-latitudes generally indicate latitude ranges of $\left|\delta\right| < 30^{\circ}$, $30^{\circ} \leq \left|\delta\right| \leq 60^{\circ}$, and $\left|\delta\right| > 60^{\circ}$, respectively. The first DI analysis of PW And was carried out by \citet{Strassmeier2006} using data from 2004. Their study revealed that cool spots were primarily distributed within an equatorial band extending up to $\pm$40$^{\circ}$ from the stellar equator. The temperature contrast between the spots and the surrounding photosphere was found to be up to 1200 K, indicating a spot temperature of 3800 K. A following DI analysis by \citet{Gu2010}, based on 2005 observations, found a different spot distribution. Their maps showed that the spots were more concentrated at intermediate to high latitudes, while weaker low-latitude spots were still present. The most recent DI analysis of PW And, conducted by \citet{Bahar2024} with spectra obtained between 2015 and 2018, simultaneously applied Doppler imaging to atomic lines and Titanium Oxide (TiO) band profiles, using high-resolution spectra, to obtain more reliable surface reconstructions. Their results confirmed the presence of a dominant high-latitude spot, with additional spots extending down to +30$^{\circ}$ latitude, consistent with the findings of \citet{Gu2010}. This suggests that a high-latitude spot has persisted for at least three years.

Young, rapidly rotating stars like PW And serve as key targets for understanding stellar magnetic activity and its role in stellar evolution. While DI has been instrumental in mapping their surfaces, all previous DI studies of PW And mentioned above have relied exclusively on spectral line profiles, without incorporating photometric constraints. This indicates the inherent limitations of using only line profiles in the reconstructions (hereafter DI-only), particularly when constraining spot latitudes and filling factors when phase coverage is incomplete or signal-to-noise ratio (S/N) is low. Simultaneous analysis of spectroscopic and photometric data remains largely unexplored, but the few existing studies indicate that these two diagnostics are complementary \citep[]{Waite2011,Finociety2021,Finociety2023}. The simultaneous analysis of DI and LCI presents a promising method to refine surface reconstructions and gain deeper insights into stellar magnetic activity. Particularly using high-precision and long-term photometric monitoring from missions like TESS, future studies can further improve our understanding of starspot evolution and magnetic field dynamics in young active stars.

In this paper, we combine newly obtained, high-resolution spectroscopic data from the 3.8 m Seimei telescope with TESS photometric observations to perform simultaneous Doppler imaging and light curve inversion (DI+LCI) to reconstruct the surface distribution of starspots on PW And. This combined dataset is uniquely powerful, as the TESS observations provide not only the photometric constraints for our spot model but also capture simultaneous flare information. By analyzing both the spectroscopic line profiles and the photometric variability, we aim to assess the improvements in spot latitude determination, the compensation for phase coverage gaps, and the overall accuracy of the reconstructed surface maps. This study provides insights into the effectiveness of the DI+LCI approach in stellar activity research and its implications for understanding the magnetic properties of young active stars.
The remainder of this paper is organized as follows.
The observations and data reductions are described in Section \ref{sec:obsdata}.
The analysis of DI and LCI is described in Section \ref{sec:3}.
The result of the analysis is shown in Section \ref{sec:results}. The result is discussed in Section \ref{sec:discuss}.
The conclusion and future prospects are described in Section \ref{sec:conclude}.

%--------------------------------------------------------------------
\section{Observations and data reduction}\label{sec:obsdata}
\subsection{High-resolution spectroscopy with Seimei/GAOES-RV}\label{ssec:specobs}
Spectroscopic observations of PW And were conducted with the Gunma Astronomical Observatory Echelle Spectrograph for Radial Velocimetry \citep[GAOES-RV;][]{Sato2024} attached to the 3.8 m Seimei Telescope \citep{Kurita2020} at Okayama Observatory of Kyoto University between 11-13 and 23-24 October in 2024. 
The average spectral resolution is \textit{R} $\sim$ 65000 with a wavelength coverage between 5160 and 5930 \AA.
The reduction of the spectra was performed using the automatic pipeline of the spectrograph (overscan subtraction, cosmic-ray removal, flat fielding, scattered-light subtraction, wavelength calibration, extraction of 1D spectrum). The slit length direction of the GAOES-RV spectrum is not parallel to the line of CCD pixels. Therefore, it is necessary to add up after performing wavelength calibration by tracing one pixel at a time in the dispersion direction.\footnote{All related IRAF cl-scripts can be downloaded via the following github page, \url{https://github.com/chimari/hds_iraf}.}
The signal-to-noise ratios (S$/$N) of the spectra were yielded $\sim$ 100-150 per pixel at 5500 \AA\ within an exposure time of 1200 seconds. Additionally, we integrated three consecutive frames to enhance S$/$N values with the effective exposure time of 3600 seconds. There are two data sets in the observing runs, which cover approximately 1.32 and 0.69 rotational cycles, respectively (Table~\ref{tab:speclog}). Since the time span between the first and second datasets corresponds to $\sim 5.5$ rotational cycles potentially introducing artifacts due to spot emerge/decay as well as spot migration, and the latter dataset exhibits insufficient phase coverage for DI, we considered only the first dataset during the analysis. However, despite the limited phase coverage of the second spectral dataset, a simultaneous reconstruction was also performed for this dataset, since TESS light curves corresponding to its observing period are available. The resulting reconstruction is provided in Sect.~\ref{sssec:second}, Fig~\ref{fig:mollmap_lsd_set-2}.

We applied the multi-line technique known as Least-Squares Deconvolution (LSD; \cite{Donati1997}) to extract high-S/N mean line profiles, thereby significantly enhancing the quality of the resulting surface brightness maps. The construction of the LSD requires a detailed line list that includes line centers and relative strengths. For this purpose, we retrieved a line list from the Vienna Atomic Line Database (VALD; \cite{Kupka1999}). To avoid introducing artifacts into the LSD profiles, we excluded wavelength regions containing spectral lines affected by chromospheric activity such as the Na I D1 and D2. The velocity step was set to 2.3 km~s$^{-1}$. The initial line mask contained several thousand atomic lines, among which 2168 could be effectively used in the LSD model. However, a significant fraction of these lines are relatively shallow (with depths just above 1\% of the continuum), which limits the S/N enhancement expected from line averaging. Consequently, the resulting LSD profiles reach a mean S/N of about 420. This rather modest improvement compared to the raw spectra (S/N=110–237) is most likely due to the narrow spectral coverage (i.e. $\sim 770$ \AA) combined with the predominance of weak lines within this range.

\subsection{TESS photometry}\label{ssec:tessflare}
The Transiting Exoplanet Survey Satellite \citep[TESS;][]{Ricker2015} observed PW And with the twenty-second cadence in the Sector 84 (October 1 to 26 in 2024).
The spectroscopic observations were contemporaneously conducted to enable combined DI and LCI analysis. During the simultaneous solution, The SAP (Simple Aperture Photometry) flux data were extracted from the Mikulski Archive for Space Telescopes (MAST).
We note that the PDCSAP (Pre-search Data Conditioning Simple Aperture Photometry) pipeline discarded the data that correspond to the time interval of the spectral observations.

\begin{table}
  \caption{Spectroscopic Observation of PW And with GAOES-RV}{%
   \setlength{\tabcolsep}{5pt} 
  \begin{tabular}{ccccccc}
      \hline
      Date & Exp. Time  & $\mathrm{BJD}_{\mathrm{Mid}}$ & $\mathrm{Phase}_{\mathrm{Mid}}$ & S/N  \\
      &  (sec.) &  &  &  \\
      \hline
      \multicolumn{5}{c}{First dataset} \\
      \hline
      11.10.2024 & 3600 & 2460594.946963 & 0.797 & 144 \\
      11.10.2024 & 3600 & 2460594.990979 & 0.822 & 185 \\
      11.10.2024 & 3600 & 2460595.040354 & 0.851 & 234 \\
      11.10.2024 & 3600 & 2460595.088444 & 0.878 & 218 \\
      12.10.2024 & 3600 & 2460595.946068 & 0.366 & 185 \\
      12.10.2024 & 3600 & 2460595.993556 & 0.393 & 207 \\
      12.10.2024 & 3600 & 2460596.040605 & 0.420 & 208 \\
      12.10.2024 & 3600 & 2460596.087699 & 0.447 & 208 \\
      13.10.2024 & 3600 & 2460596.934188 & 0.929 & 237 \\
      13.10.2024 & 3600 & 2460596.980438 & 0.955 & 155 \\
      13.10.2024 & 3600 & 2460597.030021 & 0.983 & 143 \\
      13.10.2024 & 3600 & 2460597.077324 & 0.010 & 167 \\
      13.10.2024 & 3600 & 2460597.125240 & 0.037 & 190 \\
      13.10.2024 & 3600 & 2460597.171814 & 0.064 & 186 \\
      13.10.2024 & 3600 & 2460597.217936 & 0.090 & 110 \\
      13.10.2024 & 3600 & 2460597.264082 & 0.116 & 176 \\
      \hline
      \multicolumn{5}{c}{Second dataset} \\
      \hline
      23.10.2024 & 3600 & 2460607.012362 & 0.666 & 146 \\
      23.10.2024 & 3600 & 2460607.056226 & 0.691 & 93 \\
      23.10.2024 & 3600 & 2460607.119208 & 0.727 & 111 \\
      23.10.2024 & 3600 & 2460607.210711 & 0.779 & 62 \\
      24.10.2024 & 3600 & 2460608.216293 & 0.351 & 114 \\
      \hline
    \end{tabular}}\label{tab:speclog}
\end{table}

\begin{table}
  \centering
  \caption{Adopted Stellar Parameters}
  \begin{threeparttable} 
    \renewcommand{\arraystretch}{1.2} 
    \begin{tabular}{cc}
      \hline
      Parameter & Value   \\
      \hline
      $v\sin i$ [km s$^{-1}$] & 21.9$^{+2.8}_{-1.9}$ \textsuperscript{a} \\
      $P_{\rm{rot}}$ [day] & 1.756604 $\pm$ $0.000015$ \textsuperscript{b} \\
      $\text{T}_{0}$  [HJD] & 2453200.0 \textsuperscript{c} \\
      $M$ ($M_{\odot}$)  & 0.85 $\pm$ $0.05$ \textsuperscript{d}\\
      $R$ ($R_{\odot}$)  & 1.16$^{+0.15}_{-0.11}$ \textsuperscript{c}  \\
      $i$ [$^{\circ}$] & 46.0 $\pm$ $7$ \textsuperscript{c}\\
      $T_{\rm eff}$ [K] & 5080 $\pm$ $28$ \textsuperscript{b}\\
      $T_{\rm spot}$ [K] & 3800 \textsuperscript{c}\\
      $\log g$ & 4.40 $\pm$ $0.09$ \textsuperscript{b} \\
      $[\mathrm{Fe}/\mathrm{H}]$  & -0.14 $\pm$ $0.02$ \textsuperscript{b}  \\ 
      Microturbulence  [km s$^{-1}$] & 1.93 $\pm$ $0.09$ \textsuperscript{b} \\
      Macroturbulence  [km s$^{-1}$] & 3.25 $\pm$ $0.06$ \textsuperscript{b} \\
      \hline
    \end{tabular}

    \begin{tablenotes}
      \item[a] This study.
      \item[b] \cite{Bahar2024}
      \item[c] \cite{Strassmeier2006}
      \item[d] \cite{Folsom2016}
    \end{tablenotes}
  \end{threeparttable}
  \label{tab:starparams}
\end{table}

\section{Doppler and photometric imaging}\label{sec:3}
\subsection{Updated version of {\tt SpotDIPy}}\label{sec:3.1}

The open-source DI code, {\tt SpotDIPy}, written in the Python programming language, was previously introduced in the study by \citet{Bahar2024}. In its subsequent version, {\tt SpotDIPy} adopted a three-temperature approximation instead of the previously used two-temperature model and became capable of modeling line profiles and light curves simultaneously. The code determines the brightness distribution across the stellar surface, by modeling observed line profiles and light curves through a three-temperature approach. In this method, local line profiles and continuum intensity values are used to represent the photosphere, cool spots, and hot spots. These profiles and intensities are synthesized from model atmospheres and line lists, taking into account the temperatures of the photosphere, cool spots, and hot spots, as well as the target star's surface gravity, metallicity, and microturbulence. For each surface element of the subdivided stellar surface, the local line profiles are scaled by the corresponding continuum intensities, spot filling factors ($f_s$, defined as the fractional surface coverage of the photosphere, cool spots, and hot spots), and by multiplicative factors derived from gravity darkening coefficients, as well as the projected areas and limb darkening coefficients, both of which vary with the rotational phase. The scaled local line profiles from all visible surface elements are Doppler-shifted, summed together, and normalized to produce the model line profiles for each rotational phase. For the model light curve, only the corresponding scaled continuum intensities are used.

\subsection{Combined surface reconstruction}\label{sec:3.2}

We reconstructed the surface brightness distribution of PW And using both photometric and spectral data via {\tt {\tt SpotDIPy}}. The surface grid was selected in {\tt healpy} \citep{Zonca2019, Gorski2005} discretization mode. The local line profiles were generated based on synthetic spectra computed with the MARCS atmosphere models \citep{Gustafsson2008}, using the stellar parameters listed in Table~\ref{tab:starparams} to represent both the quiet photosphere and the spots. The atomic line lists were extracted from VALD. Continuum intensities associated with the local line profiles were computed based on stellar parameters using the ExoTiC-LD Python package \citep{Grant2024}. It should be noted that under the assumption of no hot spots on the stellar surface, the surface brightness map was constructed using a two-temperature approximation. We adopted a minimum spot temperature of 3800 K, as specified by \citet{Strassmeier2006}, to model the cool spots. 

Our assumption to assume only dark spots is justified for active early-K-type stars like PW And, where observational evidence shows that the mean brightness decreases with increasing activity \citep{Lockwood07,Montet2017,Radick18}, indicating spot-dominated variability. This is consistent with surface flux transport models where distributed facular flux undergoes enhanced cancellation at high magnetic-flux injection rates \citep{Nemec22} and recent high-resolution radiative MHD simulations showing high-contrast dark starspots in K-dwarf atmospheres \citep{Bhatia25}. At the activity level and spatial resolution of our reconstruction, contributions from localized facular brightenings -- which should decrease with the angular distance from the stellar limb -- are expected to be negligible compared to the spot contribution \citep{Norris2023}.

In the simultaneous DI and LCI reconstruction, the spectroscopic and photometric datasets were assigned different relative weights to optimize the combined solution. Following the approach of \cite{Finociety2021}, we optimize the weight of the light curve so that the RMS of the residuals with respect to light curve fit becomes 0.83 mmag, the mean photometric uncertainty of the input TESS light curve (Fig.~\ref{fig:lc_and_fit}). The resulting weight of the photometric data is 0.05, hence 0.95 for the spectroscopic data. This configuration ensured a balanced contribution from both data types in the joint inversion.

\begin{figure}
 \begin{center}
  \includegraphics[width=\linewidth]{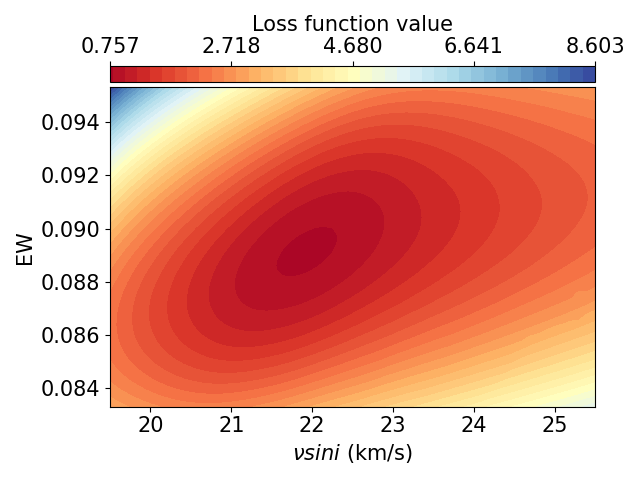}
 \end{center}
 \caption{2D grid search on the EW–$v\sin i$ plane. The colors show the loss function value.}\label{fig:grid_src}
\end{figure}

{\tt SpotDIPy} also enables fine-tuning of stellar parameters (e.g., $v\sin i$) via minimization of a loss function. As initial input for DI, we adopted the stellar parameters listed in Table~\ref{tab:starparams}. In addition to these parameters, {\tt SpotDIPy} incorporates an equivalent width (EW) parameter, which influences the depth and width of the local line profiles. The EW parameter is particularly important for matching the depth of the LSD profiles and for mitigating common DI artifacts such as spurious polar spots \citep{Cameron1994}. To optimize the EW and $v\sin i$ parameters, we performed a two-dimensional grid search in the $EW$ - $v\sin i$ parameter space, utilizing {\tt SpotDIPy}’s loss function minimization routine. The resulting loss function value distribution is presented as contour plots in Fig.~\ref{fig:grid_src}. The global loss function minimum was found at  $v\sin i = 21.9$$^{+2.7}_{-2.0}$~km~s$^{-1}$, which is consistent with the value obtained by \citet{Bahar2024}. 

\begin{figure}
 \begin{center}
  \includegraphics[width=\linewidth]{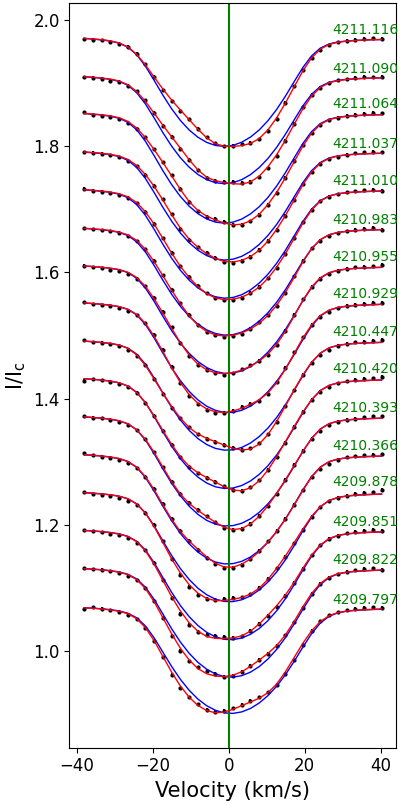}
 \end{center}
 \caption{The epoch-ordered LSD profiles derived from the observed spectra, along with their associated error bars, are represented by black circles. The epochs, marked by green text, are computed as $(\mathrm{BJD}_{\mathrm{Mid}} - \mathrm{T}_0)$ / $P_{\rm rot}$. It should be noted that the error bars are too small to extend beyond the filled circles and are therefore hard to notice. The synthetic line profiles corresponding to the spotless case are shown as blue solid lines, while the best-fit models obtained through the reconstruction process are indicated by red solid lines.}\label{fig:profs_and_fits}
\end{figure}

\begin{figure}
 \begin{center}
  \includegraphics[width=\linewidth]{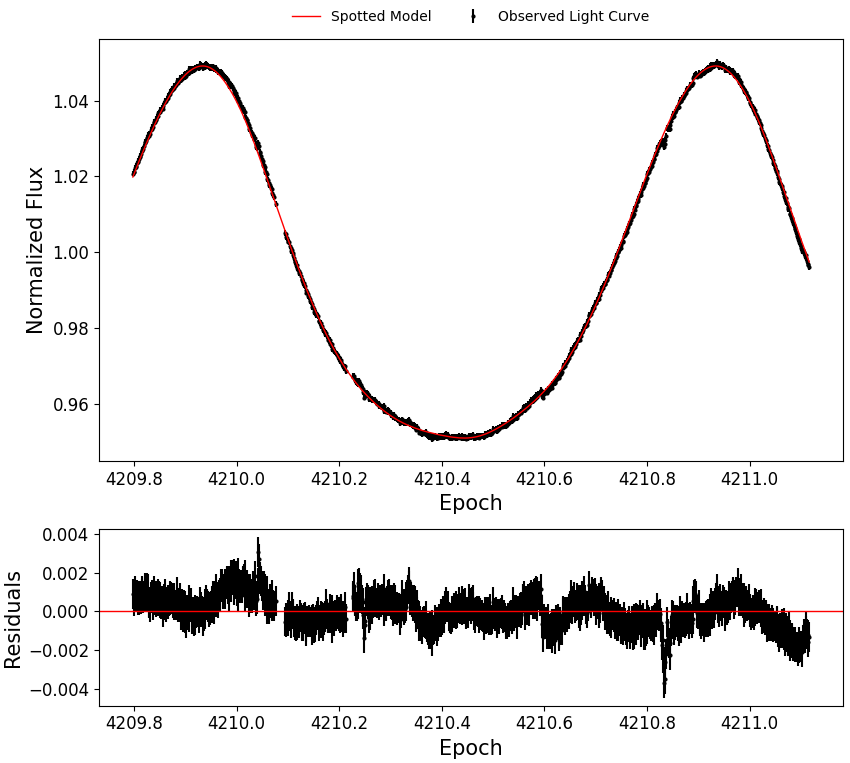}
 \end{center}
 \caption{In the upper panel, the TESS light curve as a function of epoch is plotted with black filled circles representing the data points and their associated error bars, while the red solid line denotes the best-fit model obtained from the DI+LCI reconstruction process. The lower panel displays the residuals between the observations and the model fit.}\label{fig:lc_and_fit}
\end{figure}

The LSD profiles, ordered by epoch, are shown with their best-fit models in Fig.~\ref{fig:profs_and_fits}. The light curve and its fit are shown in Fig.~\ref{fig:lc_and_fit}, plotted against the epoch. In both figures, the epochs are computed as $(\mathrm{BJD}_{\mathrm{Mid}} - \mathrm{T}_0)$ / $P_{\rm rot}$ and the fits are carried out in the combined DI+LCI mode, using the weights given above. It is evident from Fig.~\ref{fig:profs_and_fits} and Fig.~\ref{fig:lc_and_fit} that the combined DI+LCI fits are highly compatible with observed LSD profiles and light curves.

\begin{figure}
 \begin{center}
  \includegraphics[width=0.5\textwidth]{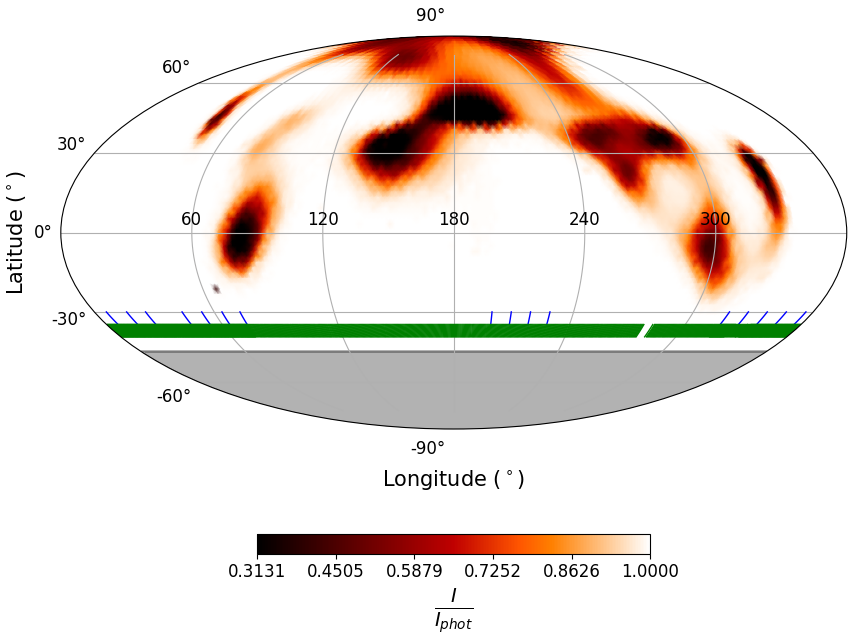}
  \includegraphics[width=0.5\textwidth]{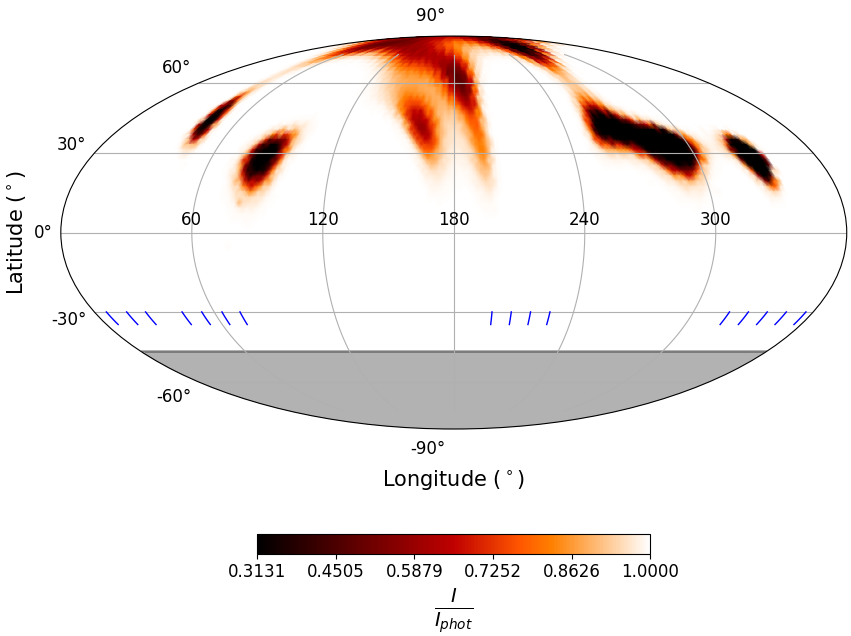}
  \includegraphics[width=0.5\textwidth]
  {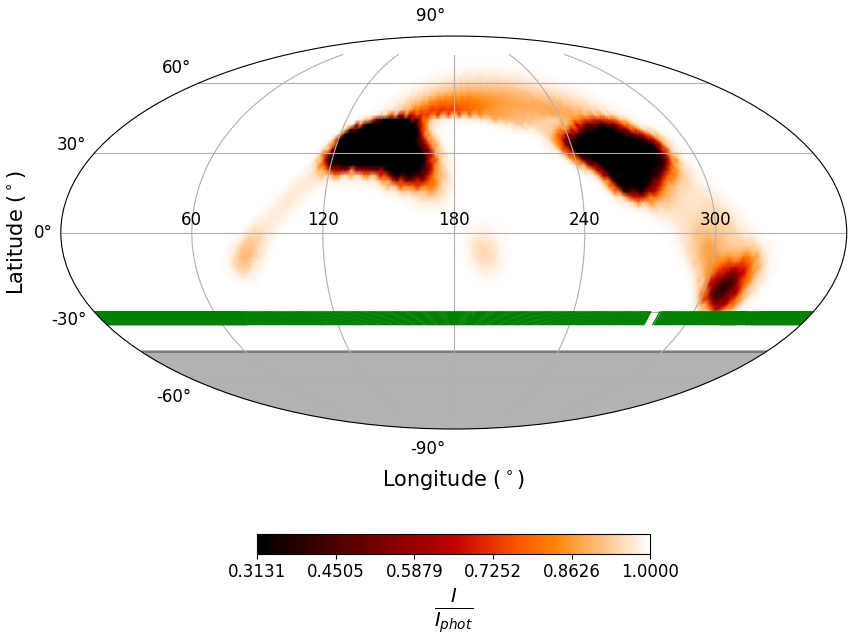}
 \end{center}
 \caption{Surface brightness distribution maps in Mollweide projection obtained from the simultaneous reconstruction of the observed light curve and LSD profiles (top panel), from LSD profiles only (middle panel), and from the light curve only (bottom panel). Blue and green tick marks indicate the longitudinal positions covered by the spectral and photometric data, respectively, shown in terms of rotational phase ($\ell = 360 \times (1 - \phi)$). The gray-shaded region indicates the invisible part of the southern hemisphere owing to the axial inclination.}\label{fig:mollmap_lsd}
\end{figure}

The resulting surface reconstruction of the combined spectroscopic and photometric (DI+LCI) solution is shown in Fig.~\ref{fig:mollmap_lsd}, in comparison with the purely spectroscopic (DI-only) and purely photometric (LCI) reconstructions. The surface map reconstructed from the simultaneous solution reveals that the spots are predominantly concentrated at mid- to high-latitudes (+30$^{\circ}$ to +90$^{\circ}$) and with additional features near and slightly below the equator, resulting in an estimated spot coverage of about 9.9\% of the visible stellar surface. A detailed discussion on the spot distribution is given in Section \ref{sec:results}.

\subsection{Flare detection} \label{sec:flare}

We employed the open-source Python package {\tt ALTAIPONY} \citep{Ilin2021} to detect stellar flares and estimate their parameters from the TESS SAP flux light curves. Prior to flare detection, we applied a custom detrending procedure to remove long-term astrophysical modulations, primarily due to starspot activity, and instrumental systematics. This involved fitting a third-order spline to nongap segments of the light curves and subtracting the fit. In addition, prominent sinusoidal variations were iteratively removed. The iterative process begins by masking outliers via $\sigma$-clipping. Isolated points above 3$\sigma$ are flagged as pure outliers, while consecutive points above 3$\sigma$ are identified as potential flares. In each iteration, an LS periodogram is computed, and a cosine fit based on the dominant frequency (from least-squares fitting) is subtracted from the light curve. This continues until the dominant peak’s S/N drops below 1.

Residual low-amplitude variability was further suppressed using a combination of Savitzky–Golay filters with window lengths of 6 and 3 hours. A cubic spline with a coarseness of 4 hours was used for baseline fitting, and a 3$\sigma$ threshold was applied to exclude outliers during the detrending. This methodology closely follows the approach outlined by \citet{Ilin2022}.

\begin{figure*}
 \begin{center}
  \includegraphics[width=18cm]{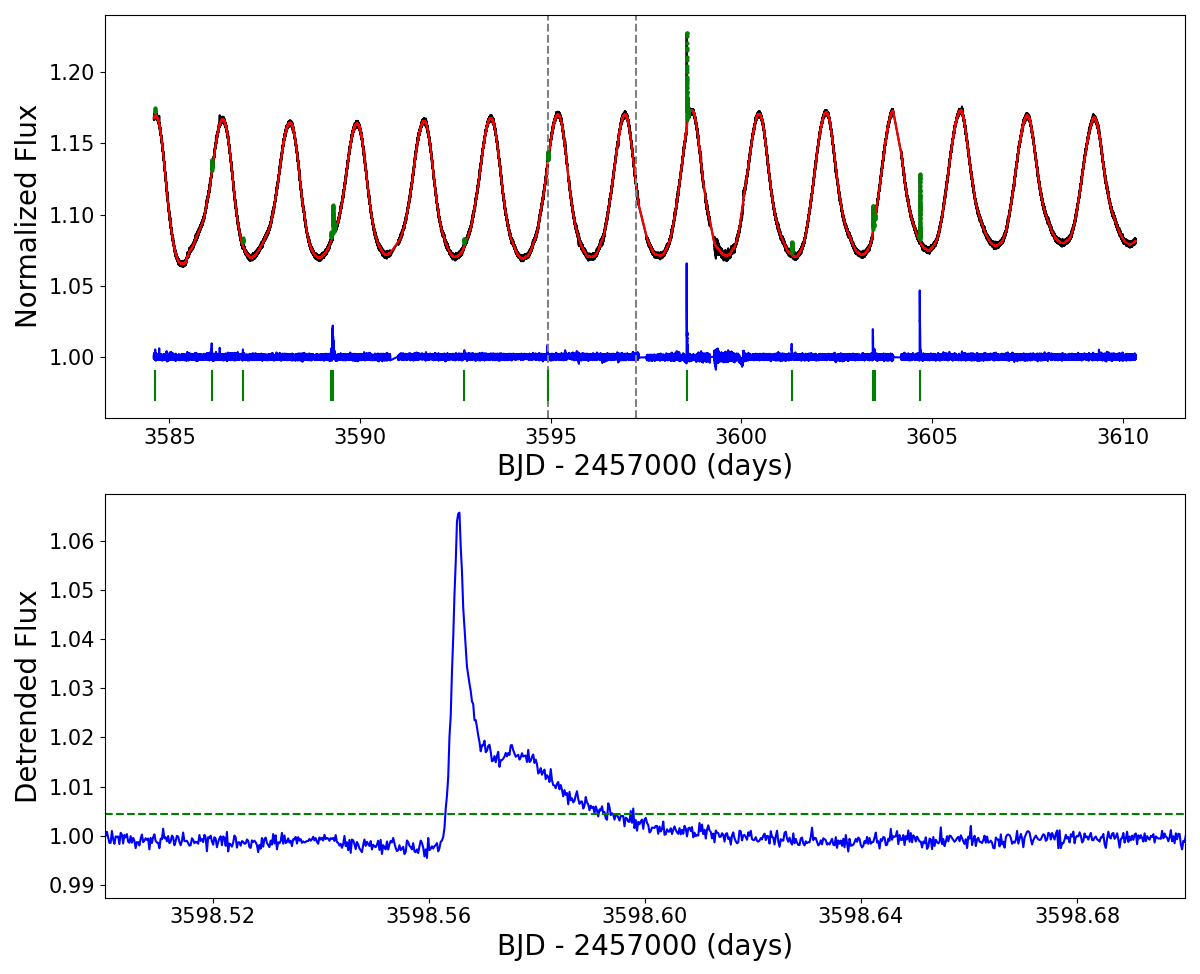}
 \end{center}
 \caption{The upper panel shows the normalized TESS SAP flux light curve of PW And from Sector 84 (October 1–26, 2024). Detrended light curve is shown as the blue line, where flare signals remain, but spot-induced modulation and instrumental effects have been filtered out. The red line denotes the model used for this detrending. Gray dashed vertical lines indicate GAOES-RV observations contemporaneous with the TESS monitoring. Flares, defined as events in which at least four consecutive data points exceed the 3$\sigma$ threshold, are marked by green points. Each detected flare is also indicated by a vertical green bar beneath the light curve. The lower panel shows an example of a detected flare, with the green dashed line representing the 3$\sigma$ threshold.}\label{fig:tesslc}
\end{figure*}

Flare detection was performed using the {\tt FlareLightCurve.find\_flares()} function of the Python package {\tt ALTAIPONY}, specifically designed for identifying stellar flares in light curve data. This function flags flare candidates based on a sigma-clipping approach applied to the detrended light curve. We adopted the parameters N1 = 3, N2 = 3, and N3 = 4, which control the detection thresholds. Flares were defined as events comprising at least four consecutive data points exceeding a 3$\sigma$ threshold. All flare candidates were visually validated, and only those exhibiting a characteristic fast rise and slower decay profile were retained. A total of twelve flares were identified (Table~\ref{tab:flareact}). The final detrended TESS light curve of PW And, the detected flares, and an illustrative flare example are presented in Fig.~\ref{fig:tesslc}.

The flare energy $E_{\mathrm{flare}}$ is calculated from the relative flux ($\Delta{F}$/F) of the TESS light curve \citep{Shibayama2013}:

\begin{equation}
E_{\mathrm{flare}} = \sigma_{\mathrm{SB}} T_{\mathrm{flare}}^4 \times  \pi R_{\mathrm{star}}^2 \times \frac{ \int R(\lambda) B(\lambda, T_{\mathrm{eff}}) \, d\lambda }{ \int R(\lambda) B(\lambda, T_{\mathrm{flare}}) \, d\lambda } \int \frac{\Delta F(t)}{F} \, dt,
\label{eq:3}
\end{equation}
where $\sigma_{\mathrm{SB}} = 5.67 \times 10^{-1} \, (\mathrm{erg} \, \mathrm{K}^{-4} \, \mathrm{m}^{-2} \, \mathrm{s}^{-1})$, $T_{\rm flare}$, $R_{\rm star}$, $T_{\rm eff}$, $R(\lambda)$, and $B(\lambda,T)$ are the Stefan–Boltzmann constant, flare temperature, stellar radius, and stellar effective temperature, response function, and Planck function under a certain wavelength $\lambda$ and temperature $T$, respectively.
The flare is assumed to be blackbody radiation with a temperature of $T_{\rm flare} = 10,000$~K \citep[]{Mochnacki1980,Hawley1992,Kowalski2013}.

\section{Results}\label{sec:results}

\subsection{The effect of photometric data on Doppler imaging}
\label{ssec:DI+LC}

The resultant spot distribution of PW And obtained in this study using the combined DI+LCI technique (Fig.~\ref{fig:mollmap_lsd}, top panel) differs from the surface brightness maps reconstructed by \cite{Bahar2024}, which also show spot concentrations at mid- to high-latitudes, but do not display any distinct features at low latitudes or the southern hemisphere. This discrepancy likely arises from methodological differences, as our analysis here employs simultaneous modeling of line profiles and light curves (obtained during the same observational epoch), while \citet{Bahar2024} relied solely on spectral profiles due to the absence of contemporaneous photometric data. Utilizing high-precision TESS photometry with full phase coverage, the reconstruction in Fig.~\ref{fig:mollmap_lsd} (top panel) shows more pronounced low-latitude features, effectively increasing the sensitivity of inversions on equatorial structures and thereby yielding a possibly more accurate representation of the surface brightness distribution. In order to highlight this difference, we also performed spot reconstruction using only LSD profiles (DI-only analysis) and then compared the results with those obtained from simultaneous DI and light curve inversion (DI+LCI analysis). 

The surface reconstruction based solely on photometric data (Fig.~\ref{fig:mollmap_lsd}, bottom panel) yields significantly less surface detail through mid- to high latitudes, indicating that most of the spatial information there is constrained by spectral line profile variability. LCI reconstructs spotted regions from about $-30^\circ$ to $+60^\circ$ latitudes. The distinguishing feature of this latitudinal band is that it encompasses regions that contributes significantly to the rotational modulation of stellar brightness owing to the stellar inclination. The rotational modulation is partly sensitive to the visibility time of a given spotted region, hence on its latitude. Incorporating photometric data improves the DI reconstruction of low-latitude spots on both sides of the equator (see Sect.~\ref{ssec:dilc_sims}). The DI-only reconstruction shown in Fig.~\ref{fig:mollmap_lsd} (middle panel) reveals a prominent polar spot near +80$^{\circ}$ - +90$^{\circ}$ latitude. Several additional spots at high latitudes are also recovered, while low-latitude and southern hemisphere spots are largely absent. Furthermore, the “equatorial” spot around $85^\circ$ longitude on the DI+LCI map appears near +30$^{\circ}$ latitude, which may be attributed to the inherent tendency of the DI technique to prioritize features in the northern hemisphere, further influenced by the sparsity of spectroscopic sampling.

These results show that the differing latitudinal sensitivities of DI and LCI are somewhat complementary. Indeed, the DI+LCI reconstruction (Fig.~\ref{fig:mollmap_lsd}, top panel) reveals significant differences in the spot distribution relative to the DI-only solution. Additional spots appear at mid-to-low latitudes, particularly around -20$^{\circ}$ - +30$^{\circ}$, including the equator. Notably, spots in the southern hemisphere become visible, which were entirely missing in the DI-only reconstruction. A pronounced spotted region between 260$^{\circ}$ - 300$^{\circ}$ longitudes in the DI-only map turns into a more structured cluster of spot regions around 20$^{\circ}$ - 40$^{\circ}$ latitudes and one new feature around $-10^\circ$ latitude. This supports the possibility that apparently monolithic round features in Doppler images might be conglomerates of smaller spot groups, similar to nests of sunspot groups. The inclusion of the light curve in addition to the line profiles enables a more reliable reconstruction of the latitudinal spot coverage, revealing detailed longitudinal patterns. 

One noteworthy feature on the DI+LCI map is the apparently monolithic spot region located between 80$^{\circ}$ - 100$^\circ$ longitudes centered at the equator. Its large size is similar to the mid-latitude one on the DI-only map mentioned above, also manifesting substantial latitudinal elongation. The simulations carried out in Sect.~\ref{ssec:dilc_sims} suggest that such an elongated feature can result from two effects: (1) blending of two smaller spot regions nearly symmetric about the equator, (2) the location of the spot being close to a wide phase gap. 

The effects of enhanced longitudinal sampling by photometry is clearly demonstrated in Fig.~\ref{fig:lat_fs}, showing the latitude distribution of the longitudinally averaged spot filling factor. The combined solution is clearly more structured than the pure Doppler image. However, it should be noted that the cross-equatorial spot occurrence is likely affected by higher-latitude features on the less visible hemisphere, owing to the rather low axial inclination \citep{Senavci2021}. 

\begin{figure}[h]
 \begin{center}
 \includegraphics[width=0.45\textwidth]{figure/lines_and_lc_map_moll.png}
  \includegraphics[width=0.45\textwidth]{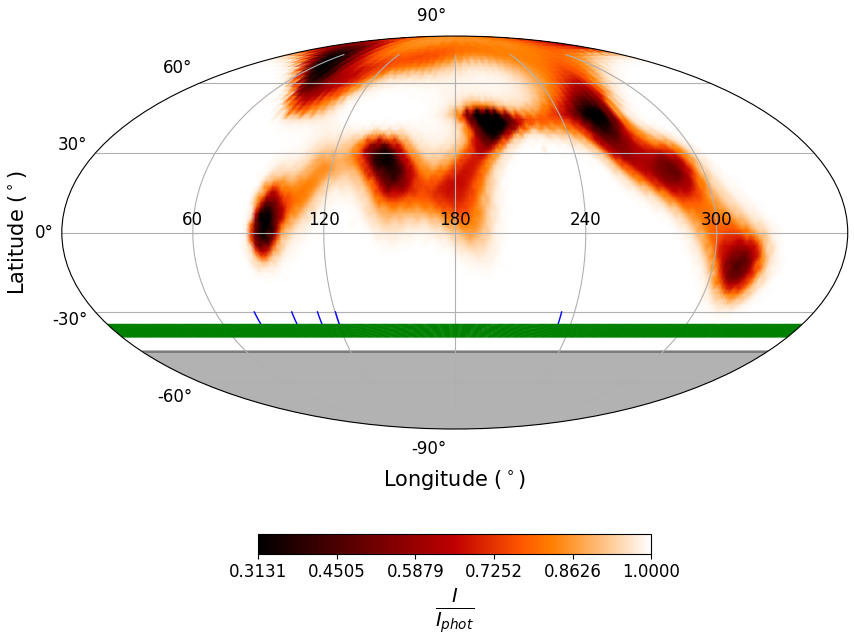}
    \par\medskip
 \end{center}
 \caption{Surface brightness distribution maps in Mollweide projection obtained from the simultaneous reconstruction of the observed light curve and LSD profiles for the first (upper panel) and second (lower panel) dataset. For description of tick marks, we refer Fig.~\ref{fig:mollmap_lsd}.}\label{fig:mollmap_lsd_set-2}
\end{figure}

\subsubsection{Simultaneous Reconstruction of Second Dataset}
\label{sssec:second}

As stated in Sect.~\ref{ssec:specobs}, a second spectroscopic dataset was obtained approximately 5.5 rotational cycles (about 10 days) after the first set. Although this dataset suffers from poor spectral phase coverage, it coincides with a period during which TESS photometric data are available. Contrary to common practice in the literature, we do not include these spectra in one common solution spread over 10 days, because the light curve morphology undergoes remarkable changes in this period (see Fig.~\ref{fig:mapvsflare}, top panel). We thus performed simultaneous DI+LCI reconstruction for this dataset as well. The phase range of the spectroscopic data is only 0.69, meaning that it does not even span a full rotation. In the simultaneous inversion, however, we used the full rotational cycle of TESS data that temporally encompasses the spectroscopic observations (specifically, the one-cycle interval preceding the epoch of the last spectrum). 

The resulting surface map is presented in Fig.~\ref{fig:mollmap_lsd_set-2}. When compared with the map derived from the first dataset, the overall spot distribution remains broadly consistent, though noticeable differences exist in spot sizes and contrasts. On the one hand, the poor phase coverage makes the reconstruction from this dataset alone unreliable; on the other hand, the resemblance of the spot distribution to that of first dataset still demonstrates the power of the combined DI+LCI inversion under sparse spectral sampling but very high photometric cadence (see Sect.~\ref{ssec:dilc_sims}).

Furthermore, considerable evolution of the spot pattern is evident. In spite of the relatively short temporal span of the two datasets, there are marked differences in the spot distribution, likely owing to spot emergence and decay on this very active star. Though the low- to mid-latitude patterns are potentially well constrained by photometry thanks to the inclination, we emphasize that the poor spectral phase coverage strongly limits the evaluation of the high-latitude regions.

\begin{figure}[t]
 \begin{center}
  \includegraphics[width=\linewidth]{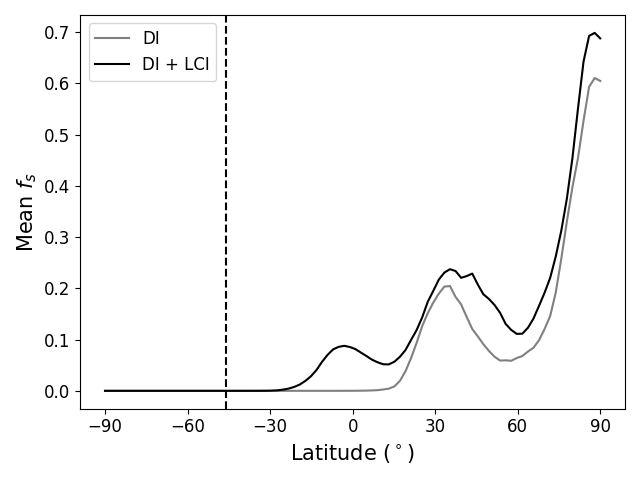}
 \end{center}
 \caption{Latitudinal distribution of mean spot coverage. The gray and black solid lines corresponds to the DI-only and the DI+LCI reconstruction, respectively. The dashed vertical line marks the latitude beyond which the stellar surface is not visible. }\label{fig:lat_fs}
\end{figure}

\subsection{Simulating the effects of photometry and spectral noise}
\label{ssec:dilc_sims}

To further assess the difference in spot latitude recovery between DI-only and DI+LCI reconstructions, we conducted simulations using the test mode of {\tt SpotDIPy}.
For each test, we produced an artificial map with a prescribed spot distribution and generated corresponding synthetic line profiles using the same local line profiles as in the DI procedure applied to PW And, as well as synthetic light curves, all computed using the stellar parameters of PW And. The total spotted area was kept constant in all cases.

\subsubsection{Spectra with ideal phase coverage under varying S/N cases}
\begin{figure*}
 \begin{center}
 \setlength{\unitlength}{1cm}
 \begin{picture}(18,9)
   % Sol panel (S/N = 2000)
   \put(0,0){\includegraphics[width=0.45\textwidth]{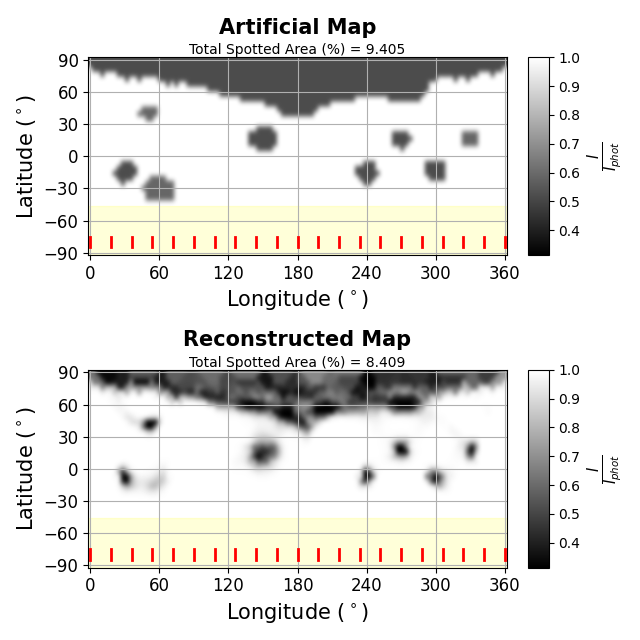}}
   \put(8.4,8.2){\makebox(0,0){S/N = 2000}}
   % Sağ panel (S/N = 420)
   \put(9.3,0){\includegraphics[width=0.45\textwidth]{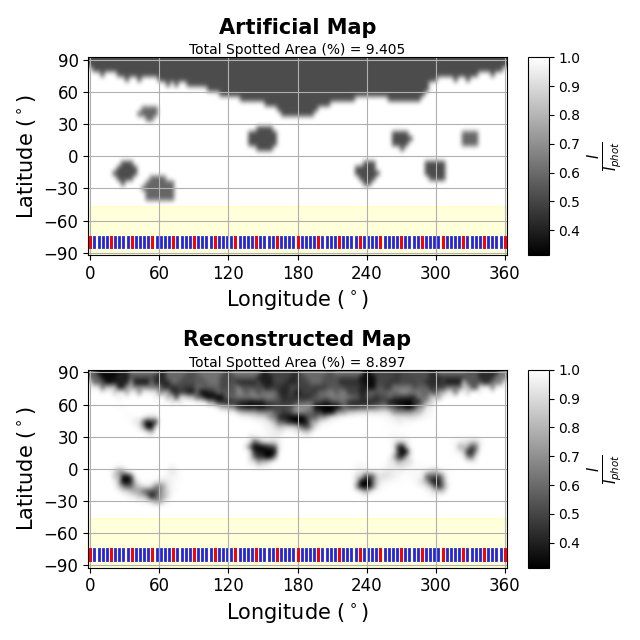}}
 \end{picture}
 \end{center}
 \caption{Reconstructed spot distribution maps obtained from DI-only (left panels) and combined DI+LCI (right panels) inversions (lower panels), based on the artificial input map (upper panels) under high S/N conditions and uniform phase sampling. The yellow-shaded region indicates the invisible part of the southern hemisphere, owing to the axial inclination. The red and blue tick marks show the phase coverage of spectral and photometric data, respectively.}
 \label{fig:ideal_high_snr}
\end{figure*}

% Ideal case: high S/N, dense phase sampling
We begin with an \emph{ideal-case} simulation, where we set a uniform phase sampling interval of 0.05 and added noise to the produced LSD profiles, so that their S/N is 2000. We also generated a high-cadence synthetic light curve from the input map, with a TESS-like sampling interval. We carry out this experiment first in DI-only mode, and then in DI+LCI mode.
The input and reconstructed surface maps are shown in Fig.~\ref{fig:ideal_high_snr} for DI-only mode (left panels) and for DI+LCI (right panels). Owing to the very `ideal' conditions, the DI-only reconstruction recovered the artificial spot distribution reasonably well, reproducing both high- and low-latitude spots with a high accuracy. The only exception is the spot at $-30^\circ$ latitude. This spot is reconstructed to about $5^\circ$ northward latitude and the filling factor is highly underestimated. 
In the combined (DI+LCI) mode, the surface distribution of spots turns out to be similar to the DI-only mode. The southern spot that was failed to be reproduced in DI-only mode is recovered with much better accuracy here. The spot has a distinct profile, though it partially blends with the neighboring spot. Also, the combined mode recovered a slightly more accurate spot area coverage.

\begin{figure*}
 \begin{center}   
  \setlength{\unitlength}{1cm}
   \begin{picture}(18,9)
    \put(0,0){\includegraphics[width=0.45\textwidth]{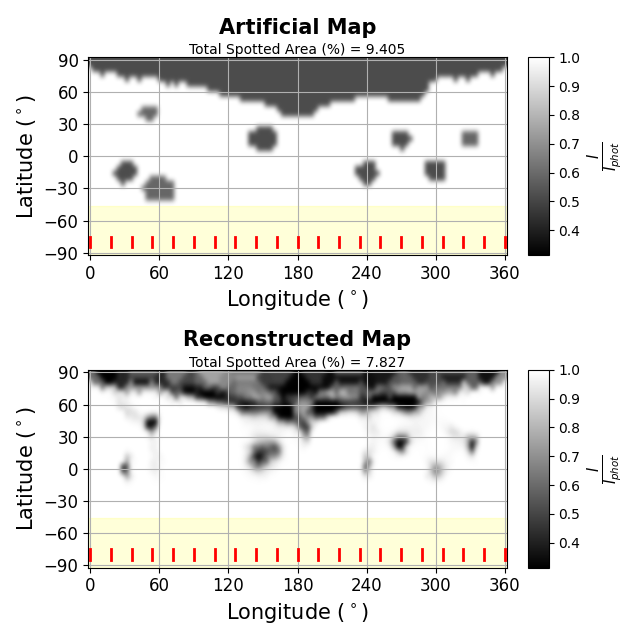}} 
    \put(8.4,8.2){\makebox(0,0){S/N = 420}}
    \put(9.3,0){\includegraphics[width=0.45\textwidth]{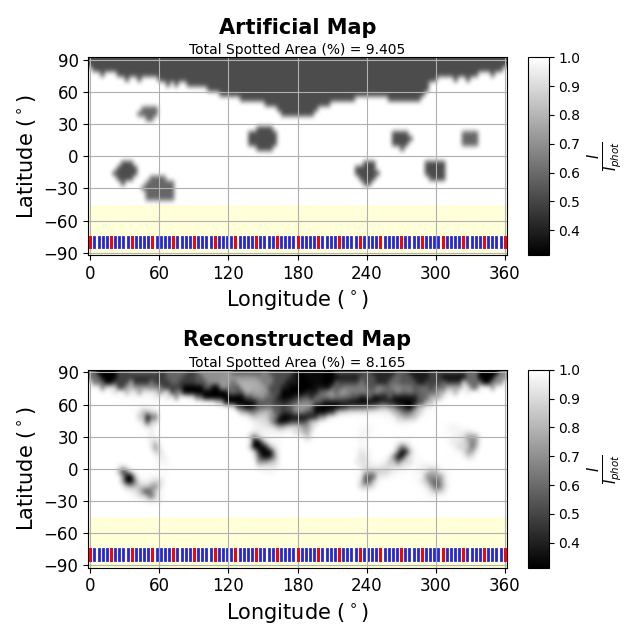}} \par\medskip
    \end{picture}
 \end{center}
 \caption{Same as Fig.~\ref{fig:ideal_high_snr}, but for low S/N.}\label{fig:ideal_low_snr}
\end{figure*}

% Noisy case: low S/N, dense phase sampling
In the low-S/N case (S/N = 420) (Fig.~\ref{fig:ideal_low_snr}), the impact of noise on DI reconstructions becomes more pronounced. Despite good phase sampling, DI-only (Fig.~\ref{fig:ideal_low_snr}, left panel) is unable to recover the low-latitude spots: The spots at $-15^\circ$ latitude appeared at the equator with underestimated filling factor, and the one at $-30^\circ$ is not reconstructed at all. The equatorward extension from the spot at $+40^\circ$ latitude is the imprint of this southern-hemisphere spot \citep[see also][Fig.~13]{Senavci2021}. 
This low fidelity is a consequence of the weak sensitivity in line-profile inversions (DI-only) to low-latitude spots, especially under low-S/N conditions.
Inaccuracies are also remarkable in the shape and contrast of spots. 

The DI+LCI reconstruction (Fig.~\ref{fig:ideal_low_snr}, right panels) is, however, more robust, recovering spot latitudes and morphologies more reliably, despite lower S/N of the LSD profiles. In contrast to the DI-only mode, the spot at $-30^\circ$ latitude is recovered well, though at a slightly lower filling factor than in the high-S/N case. It is placed at a slightly lower longitude and lower latitude than in the input map, though it is still distinguishable from the neighboring spot.
Here, the low S/N stands out as a limiting factor, leading to the underestimation of spot filling factor.

In both the DI-only and DI+LCI cases, the spot filling factors are underestimated as expected, though in the DI+LCI mode less significantly.

\subsubsection{Spectra with the observed phase coverage under varying S/N
cases}
\begin{figure*}
 \begin{center}
   \setlength{\unitlength}{1cm}
   \begin{picture}(18,9)
   \put(0,0){\includegraphics[width=0.45\textwidth]{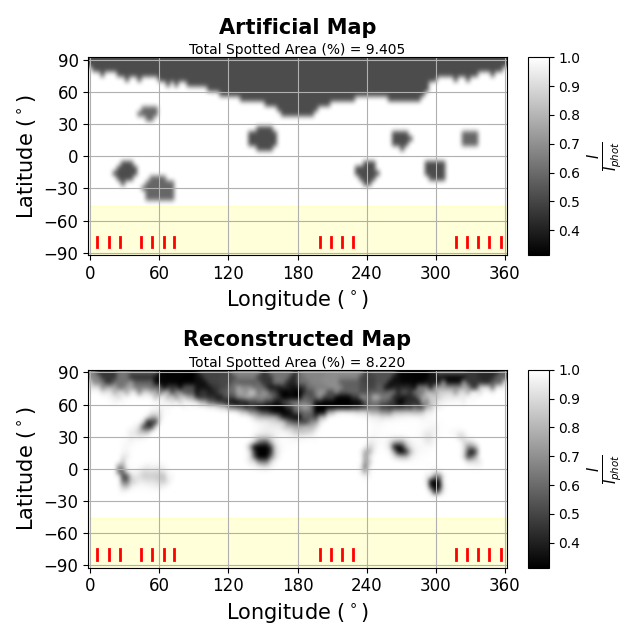}}
   \put(8.4,8.2){\makebox(0,0){S/N = 2000}}
    \put(9.3,0){\includegraphics[width=0.45\textwidth]{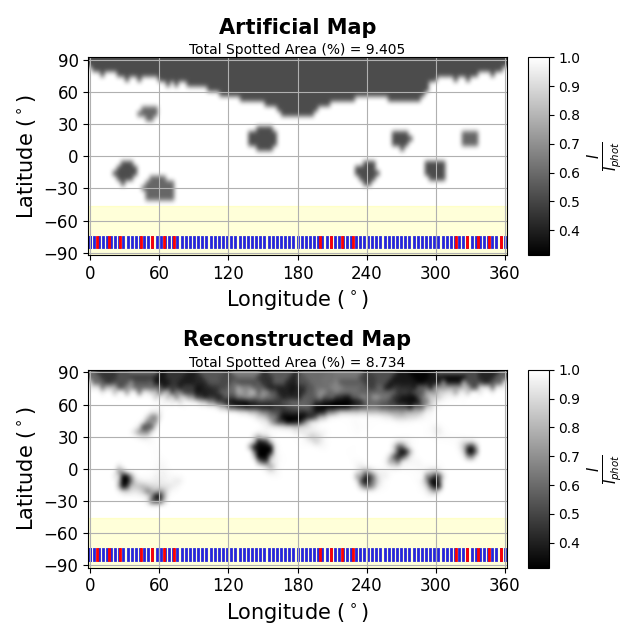}} \par\medskip
    \end{picture}
 \end{center}
  \caption{Same as Fig.~\ref{fig:ideal_high_snr}, but for spectral phase coverage of PW And.}\label{fig:low_pcov_high_snr}
\end{figure*}

\begin{figure*}
 \begin{center}
    \setlength{\unitlength}{1cm}
   \begin{picture}(18,9)
    \put(0,0){\includegraphics[width=0.45\textwidth]{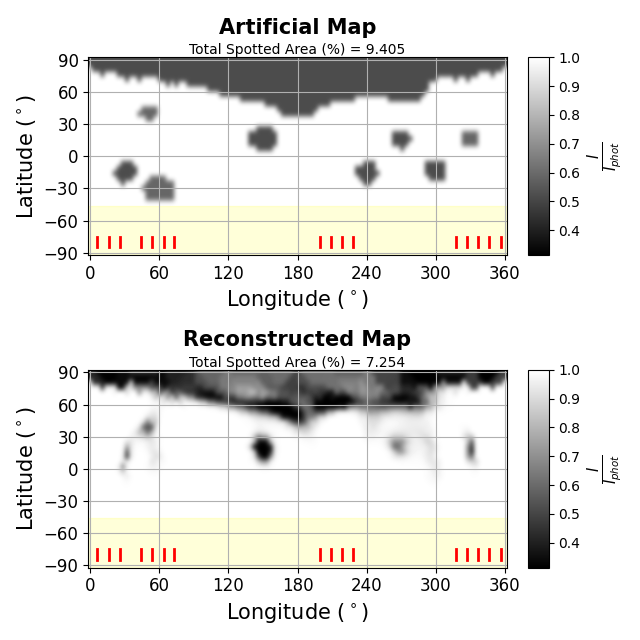}} 
    \put(8.4,8.2){\makebox(0,0){S/N = 420}}
    \put(9.3,0){\includegraphics[width=0.45\textwidth]{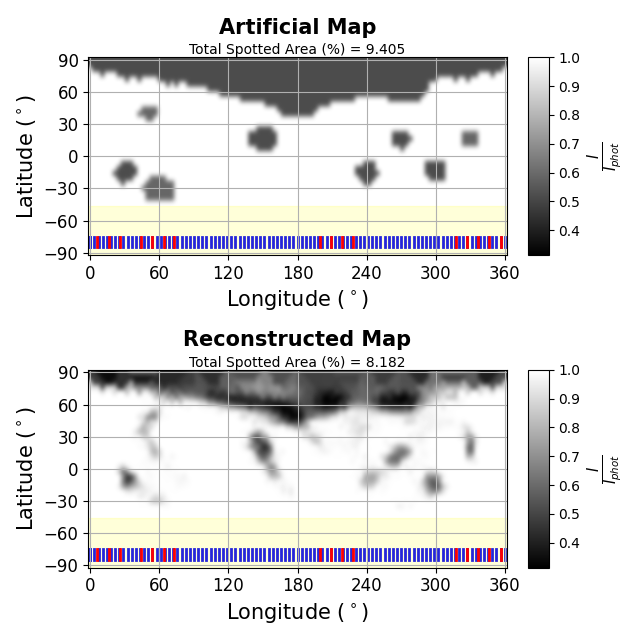}} \par\medskip
    \end{picture}
 \end{center}
  \caption{Same as Fig.~\ref{fig:ideal_high_snr}, but for spectral phase coverage and S/N of PW And.}\label{fig:low_pcov_low_snr}
\end{figure*}

To examine the impact of phase coverage, we apply the observed rotational phases of PW And for the input synthetic LSD profiles, again for S/N = 2000 and 420. The results are shown in Figs.~\ref{fig:low_pcov_high_snr}-\ref{fig:low_pcov_low_snr}. Such irregular phase gaps, where some longitudes are poorly covered is a common problem in single-spectrograph observing runs. Using these datasets, we separately assessed the impact of phase coverage and spectral noise on the accuracy of reconstructed spot maps.

In the high-S/N case (S/N = 2000, Fig.~\ref{fig:low_pcov_high_snr}), the DI-only solution gives a less accurate map than the uniformly sampled case. Except for the southern spots at 30$^{\circ}$ and $300^\circ$ longitudes, the near-equatorial performance is negatively impacted by the irregular phase coverage. When we include densely sampled photometry, however, the reconstruction performance becomes very similar to the ideal case (Fig.~\ref{fig:ideal_high_snr}) as it reproduces both high- and low-latitude spots with good accuracy.

The “realistic” low-S/N case (uneven sampling and S/N = 420, as in PW And case) is shown in Fig.~\ref{fig:low_pcov_low_snr}, where the impact of noise on DI reconstructions becomes more pronounced. In DI-only case, spots that should appear in the southern hemisphere are reconstructed at northern latitudes along the same longitude, likely due to the difficulty in distinguishing weak spot signatures under high noise conditions. For most spots, equatorward extensions are present with the spot filling factor strongly dispersed along the latitude, which is a common problem in DI-only imaging. Here, we show that this effect comes from the proximity of neighboring spots, the low S/N, and partly by phase gaps. The overall spot filling factors were also underestimated, as the increased noise level introduced additional uncertainties in the inversion process, leading to a suppression of weaker spot features. Finally, DI+LCI recovers the southern spots near their correct latitudes, producing more accurate morphology and contrast, although the accuracy is still somewhat less than the ideal case.

As an additional simulation showing the benefits of simultaneous DI+LCI reconstruction, we placed circular spots with locations, sizes, and contrasts similar to the DI+LCI map in Fig.~\ref{fig:mollmap_lsd} top panel and applied the same phase coverage and spectral S/N on the DI+LCI solution, where LCI employs the artificial light curve generated from the input map. The input spot parameters are given in Table~\ref{tab:spots}. The resulting map is shown in Fig.~\ref{fig:dilc_realistic}, being qualitatively similar to the input map. In other simulations (not displayed here), inclusion of additional southern-hemisphere spots led to their successful recovery, though shifting them toward the equator by +5$^{\circ}$ - +10$^\circ$. The two spot regions close to $90^\circ$ longitude, located nearly symmetric about the equator at $\pm 8^\circ$ latitudes was reconstructed as a blended monolithic spot with an elongation reflecting the input configuration of two spots on two hemispheres.

\subsection{Spot Latitude Effects on Light Curve Variability}\label{ssec:spotlat}
To test the sensitivity of the light curve to the latitudinal distribution of spots, we conducted an additional simulation using artificial data. We first created a realistic reference map based on the longitudinal distribution of PW And and then generated a series of synthetic light curves from modified maps in which the spot latitudes were systematically shifted (e.g., to higher latitudes or into the southern hemisphere). The key result was that these significant changes in spot latitude produced only negligible variations in the amplitude and morphology of the synthetic light curves.

\subsection{Flares and spots}\label{ssec:flares}
We investigate the correlation between flare occurrence and starspot distribution by using flares detected in the TESS light curves and starspot maps reconstructed from the combined DI and LCI solution \citep[e.g.,][]{Ikuta2023}. Although a single DI+LCI map of the full stellar surface is used, we refer to its projected hemispheric views at various flare phases as “maps” for clarity in phase-resolved comparison. A flare occurring at a given phase is assumed to originate from starspots located on the visible hemisphere. Therefore, we examine whether flare events tend to coincide with rotational phases where prominent spot features are visible in the DI+LCI maps, in order to infer a possible spatial association between flare activity and spot distribution. We detect twelve flares in the TESS light curves (Sect.~\ref{sec:flare}). For each flare, the corresponding stellar longitude is computed from the rotational phase at the flare's occurrence time, using the rotation period $P_{\rm rot}$ and reference time ${\rm T}_0$ (Table~\ref{tab:starparams}). For each flare event, we examine the distribution of starspots on the visible hemisphere as reconstructed from the DI+LCI map at the corresponding rotational phase.

\begin{figure*}
 \begin{center}
  \includegraphics[width=12cm]{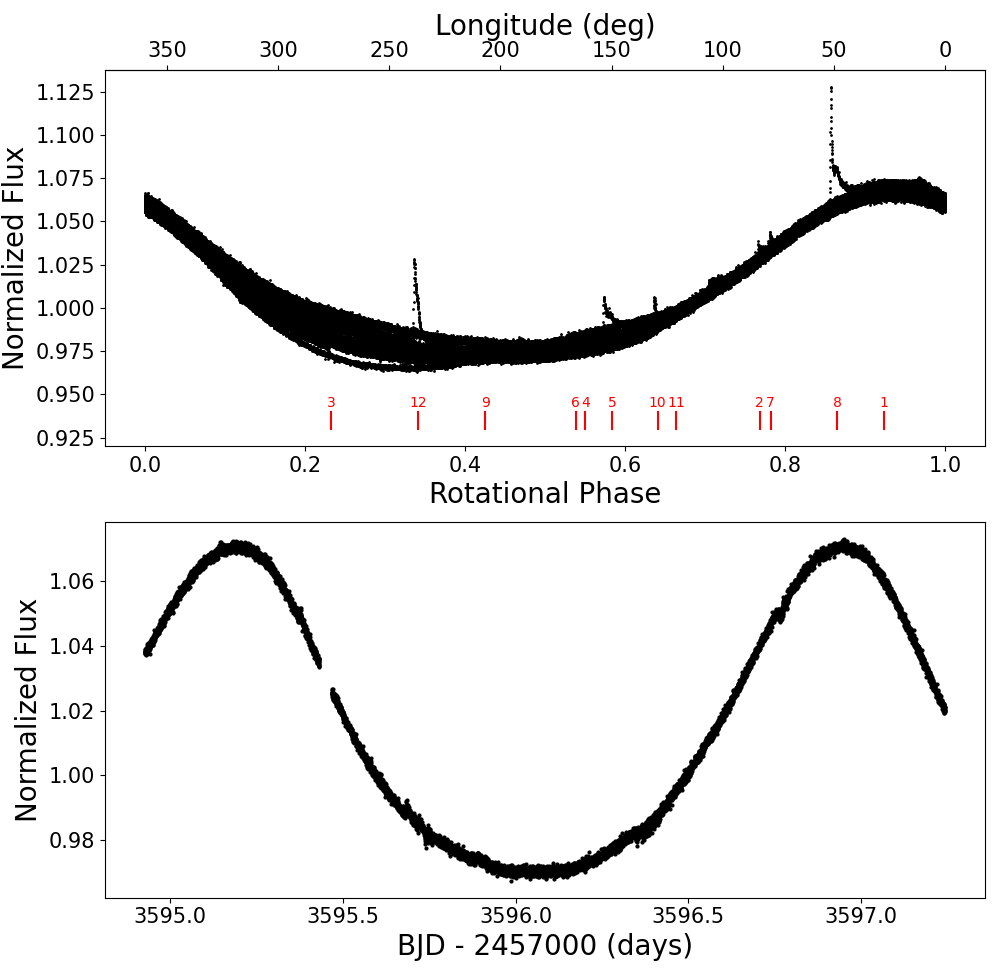}
  \hspace*{3cm}\includegraphics[width=14cm]{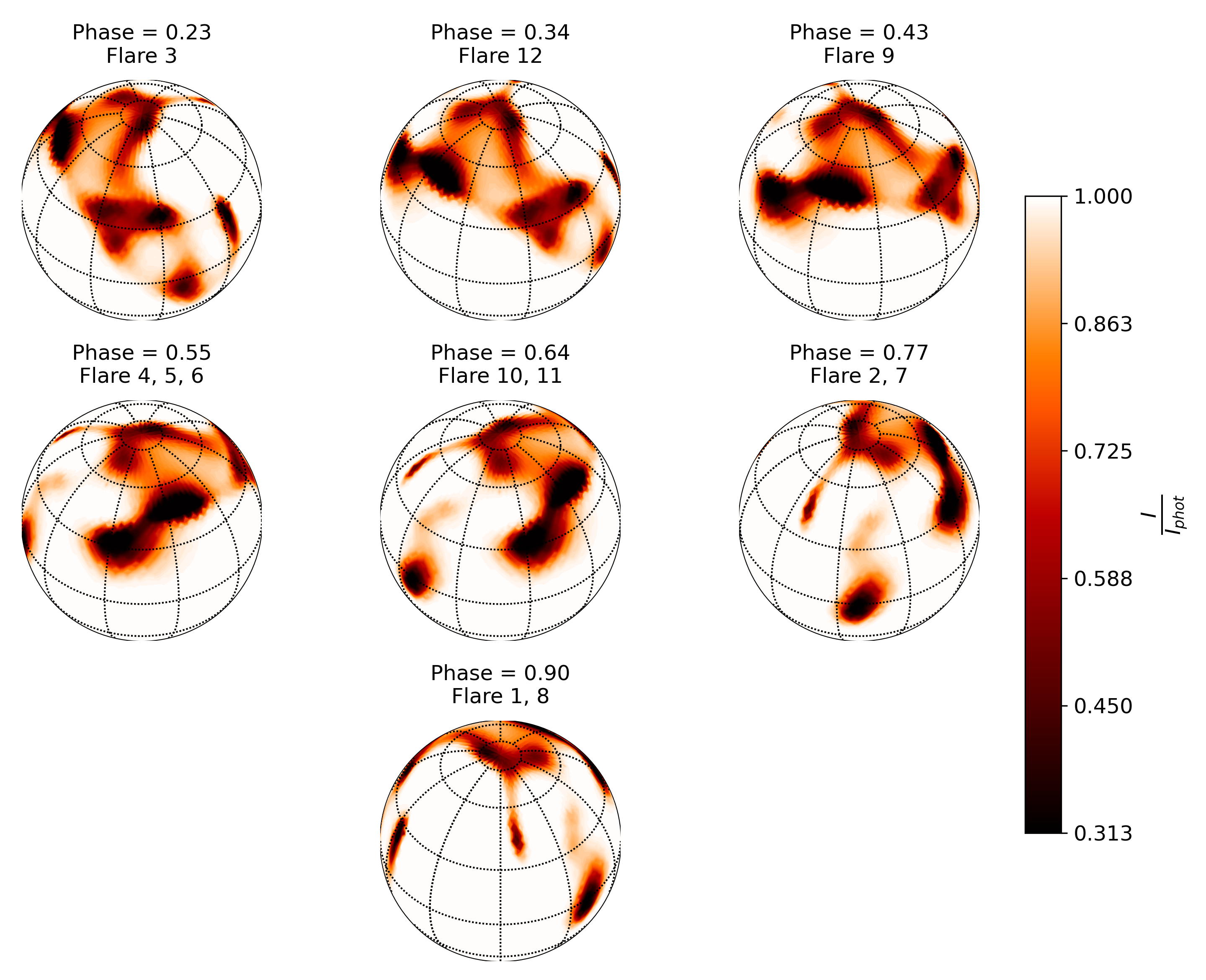}
 \end{center}
 \caption{The upper panel shows the TESS light curve of PW And from Sector 84, phase-folded over the stellar rotation period. The detected flares are marked by vertical red lines and numbered. The middle panel represents the light curve covering the same time span as the spectroscopic data. The lower panel shows the DI+LCI-reconstructed surface maps at the phases corresponding to each flare.}\label{fig:mapvsflare}
\end{figure*}

Fig.~\ref{fig:mapvsflare} displays the TESS light curve, phase-folded with the stellar rotation period, indicating the rotational phases at which flares occurred. During Sector 84, the light curve morphology exhibited some clear morphology changes, but its overall modulation amplitude remained moderately stable, suggesting that the total spotted area did not drastically change. The rotational phases surrounding each flare event are displayed to allow a direct temporal and spatial comparison between flare timing and spot distribution. The lower panel of Fig.~\ref{fig:mapvsflare} presents the corresponding DI+LCI maps reconstructed at each flare phase, enabling an approximate spatial association to be assessed. Additionally, Table~\ref{tab:flareact} summarizes the flare properties, including the flare ID, Barycentric Julian Date (BJD), rotational phase, longitude derived from rotational phase, hemispheric spot filling factor (computed with {\tt SpotDIPy}), and flare energy ($E_{\mathrm{flare}}$). The hemispheric spot filling factor represents the percentage of the visible hemisphere covered by spots at the time of each flare.

\begin{table*}
  \centering
  \caption{Detected Flares of PW And}
  \begin{threeparttable} 
    \renewcommand{\arraystretch}{1.2} 
    \begin{tabular}{ccccccc}
      \hline
      Flare  & BJD & Rotational phase & Longitude (deg) \textsuperscript{\textdagger} & Hemispheric spot filling factor\textsuperscript{*}  & $E_{\mathrm{flare}}$ (erg) \\
      \hline
        1 & 2460584.62965164 & 0.924 & 27 & 0.131 & $6.39\times 10^{32}$\\
        2 & 2460586.11276864 & 0.768 & 83 & 0.118 & $4.36\times 10^{33}$\\
        3 & 2460586.92944396 & 0.233 & 276 & 0.136 & $8.94\times 10^{32}$\\
        4 & 2460589.24370010 & 0.551 & 162 & 0.138 & $1.12\times 10^{33}$\\
        5 & 2460589.30295976 & 0.584 & 150 & 0.134 & $4.77\times 10^{34}$\\
        6 & 2460592.73549572 & 0.538 & 166 & 0.138 & $1.07\times 10^{33}$\\
        7 & 2460594.91998273 & 0.782 & 78 & 0.119 & $3.26\times 10^{33}$\\
        8 & 2460598.57967474 & 0.865 & 48 & 0.131 & $7.17\times 10^{34}$\\
        9 & 2460601.32038820 & 0.426 & 207 & 0.132 & $3.89\times 10^{33}$\\
        10 & 2460603.45508208 & 0.641 & 129 & 0.124 & $2.25\times 10^{34}$\\
        11 & 2460603.49524350 & 0.664 & 121 & 0.121 & $1.08\times 10^{33}$\\
        12 & 2460604.68538625 & 0.341 & 237 & 0.147 & $5.71\times 10^{34}$\\
      \hline
    \end{tabular}

    \begin{tablenotes}
      \item[\textdagger] The longitude corresponding to the rotational phase of each flare event.
      \item[*] The hemispheric spot filling factor is defined as the percentage of the total spot coverage on the hemisphere visible to the observer at a given rotational phase.
    \end{tablenotes}
  \end{threeparttable}
  \label{tab:flareact}
\end{table*}

\begin{figure*}
 \begin{center}
 \includegraphics[width=8cm]{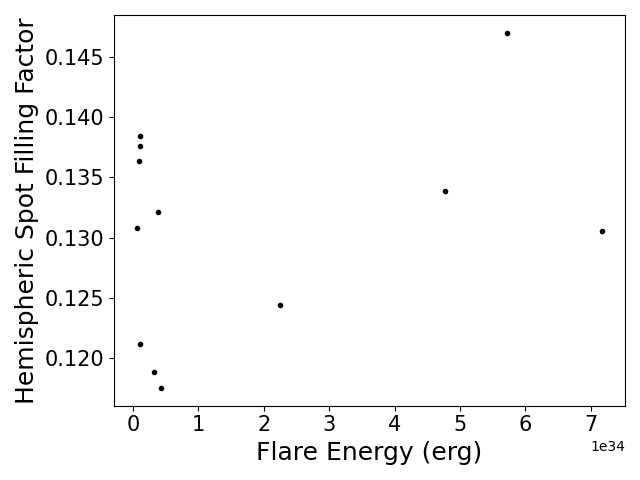}
 \end{center}
 \caption{Bolometric flare energy as a function of the hemispheric spot filling factor. The flare energies show a large scatter, spanning over an order of magnitude, and exhibit no clear correlation with the hemispheric spot filling factor.}\label{fig:fsvsflare}
\end{figure*}

The DI+LCI-derived maps were compared with the flare longitudes derived from their rotational phases to examine whether the detected flares are associated with specific starspot regions. The results (Fig.~\ref{fig:mapvsflare}) indicate that the flare events are distributed over a broad range of longitudes. This distribution appears consistent with the longitudinally dispersed spot configuration revealed by the surface map, which exhibits several moderately sized starspots across all visible longitudes, rather than one or two dominant active regions. While the spots are spread in longitude, they are primarily located at mid-to-high latitudes, approximately between +30$^{\circ}$ and +60$^{\circ}$.

Despite this potential spatial association, our analysis does not reveal a clear correlation between flare energy and local spot properties. Fig.~\ref{fig:fsvsflare}, which plots bolometric flare energy against the hemispheric spot filling factor, shows no discernible trend. Furthermore, flare energies vary by more than an order of magnitude, even among events occurring near similar active longitudes (Table~\ref{tab:flareact}).

\section{Discussion}\label{sec:discuss}
\subsection{Enhanced spot mapping with simultaneous DI+LCI}\label{ssec:simultdilc}
Our comprehensive DI analysis of PW And demonstrates the importance of simultaneously incorporating spectroscopic and photometric information when reconstructing the stellar surface spot distribution, particularly under observational constraints such as moderate axial inclination, incomplete phase coverage, and limited S/N.

As the comparison between our DI-only and DI+LCI maps reveals (Sect.~\ref{ssec:DI+LC}), the inclusion of photometric data is critical for recovering spots at low latitudes and in the southern hemisphere, which are systematically missed by DI-only methods. Controlled simulations using artificial spot maps further confirm that DI-only reconstructions consistently underestimate the spotted area, particularly in the southern hemisphere (Figs.~\ref{fig:ideal_high_snr}-\ref{fig:low_pcov_low_snr}). The limitations observed in DI-only reconstructions primarily arise from the combined effects of axial inclination, low spectral S/N, and incomplete phase coverage. The adopted inclination of +46° causes the northern hemisphere to remain preferentially and more continuously visible, inherently reducing the detectability of southern features. When coupled with lower S/N and irregular phase sampling, this geometric bias is further amplified, leading to weaker Doppler signatures and increased latitude ambiguities. Consequently, while high-latitude northern spots produce distinct and well-defined line-profile distortions, southern hemisphere features suffer from reduced visibility, stronger projection effects, and diminished radial velocity modulation, all of which limit their reliable recovery in DI-only inversions. This bias in Doppler imaging towards features that remain visible for a larger fraction of the rotation, particularly in the visible hemisphere, is well-documented \citep[e.g.,][]{Berdyugina2005}. At moderate axial inclinations like that of PW And, the ability to detect spots in the equatorial region or on the southern hemisphere is therefore significantly diminished, often leading to reduced contrast, suppressed filling factors, or even ghost-like features at incorrect latitudes, as seen in our DI-only map.

In contrast, the DI+LCI map of PW And successfully recovered features that were undetectable or mislocated in the DI-only analysis. This improvement demonstrates the compensatory role of photometric constraints in correcting for inclination-induced visibility biases and recovering surface features with limited spectroscopic signatures. This trend has also been reported for ZDI, where the technique is most sensitive to features at higher latitudes in the visible hemisphere and less effective for low-latitude spots \citep{Vogt1987,Brown1991,Finociety2021}.

We further investigated the impact of spectral S/N and phase coverage on the accuracy of surface reconstructions. The simulations clearly show that DI-only inversions are highly sensitive to these observational parameters. Under idealized conditions with high S/N (2000) and uniform phase sampling, DI-only reconstructions successfully reproduce both high- and low-latitude spots with good fidelity. However, their performance declines significantly under more realistic observing conditions characterized by lower S/N (420) and irregular phase coverage. In such cases, DI-only maps systematically underestimate spot filling factors and frequently misplace southern hemisphere spots toward higher latitudes. This is a well-known degeneracy arising from weak line-profile sensitivity to low-latitude regions.

In contrast, the combined DI+LCI inversion demonstrates greater resilience to these observational limitations. Even under incomplete phase sampling and reduced S/N, the inclusion of photometric information significantly improves the recovery of spot morphology and latitude, particularly for features located in the southern hemisphere. Small latitude offsets ($\sim 5$$^{\circ}$ - 10$^{\circ}$) and slight underestimations of filling factors persist. The main reason is the axial inclination of 46$^{\circ}$, limiting the latitudinal visibility. The overall reconstructed surface remains far more consistent with the input configuration than in DI-only cases.

These findings emphasize that incorporating both spectroscopic and photometric data provides a stronger foundation for reliable surface imaging, compensating for weakened Doppler signatures in low-visibility regions. Nonetheless, reconstruction accuracy still depends critically on optimizing phase sampling and maintaining sufficient S/N. Achieving an optimal balance between these two data types is therefore essential to minimize artifacts and ensure robust latitude recovery in Doppler imaging analyses.

Our additional tests on latitude redistribution (Sect.~\ref{ssec:spotlat}) show that shifting spots toward the southern hemisphere does not significantly alter the synthetic light curve morphology, as the inclination reduces the projected area of southern spots and minimizes their photometric impact. This experiment demonstrates that photometric data alone may provide poor constraints on spot latitude, particularly in the obscured hemisphere, confirming the well-known degeneracy in light curve inversions where different latitude distributions can produce nearly identical light curves. Therefore, while photometry is crucial for constraining low-latitude features in combination with DI, our results  highlight the indispensable role of using the Doppler effect to locate spots from line profiles.

\subsection{Latitudinal distribution of activity}
\label{ssec:latdist}
The overall latitudinal distribution of spots on this rapidly rotating young K2 star is consistent with previous studies reporting that active regions on young active G- to early-K stars often appear at mid-to-high latitudes \citep{Strassmeier2009}. The existence of low-latitude features on a fast-rotating K2V star with a larger convection-zone fraction than a G2V star should be less expected from flux-tube rise geometry, as modelled by \citet{Isik2011} for a K0V dwarf rotating with a period of 2 days, close to that of PW And (K2V, 1.76 d). This three-component model consisting of an interface dynamo, rising flux tubes, and surface flux transport led to a surface activity cycle period of about 3 years with strong maxima dominated by high-latitude features and still active minima dominated by mid-latitude spots. However, our composite DI+LCI reconstructions of PW And does show low-latitude spots within $\pm 30^\circ$ latitude, along with mid- to high-latitude ones. Under the hypothesis that active region producing flux loops stem from the lower convection zone, such a pattern requires intense flux tubes to be formed, leading to buoyancy-dominated rise and emergence at low latitudes, which was shown by \citet{Isik2024} for a solar-type star rotating at $P_{\rm rot}\sim 3$~d.

\subsection{Flare in relation to spot distribution}\label{ssec:flarespot}
Although not statistically significant, the potential correlation between flare occurrence timing and reconstructed spot longitudes (Sect.~\ref{ssec:flares}) may suggest an association between magnetic active regions and flare activity on PW And. While it remains difficult to uniquely identify the origin of each flare due to the complex and widespread spot distribution on the stellar surface, some flares were detected at rotational phases corresponding to longitudes where relatively prominent spot structures are reconstructed (Table~\ref{tab:flareact} \& Fig.~\ref{fig:mapvsflare}). These spots are primarily located at mid-to-high latitudes (+30$^{\circ}$ and +60$^{\circ}$.), which may indicate a preferred latitude range for flare-productive regions in this system. A much larger sample of flare events is required to statistically confirm this trend.

In contrast to the Sun, where flare activity often follows the butterfly diagram with a bias toward low latitudes \citep{Joshi2005}, young active stars — particularly M dwarfs — have, in some cases, been found to exhibit flare activity at higher latitudes \citep{Ilin2021b}. Whether this trend extends to K-type stars like PW And is still uncertain. Some studies suggest wider latitude distributions in G–K–M stars, possibly with a low-latitude bias \citep{Yang2025}, indicating that stellar type and age may significantly influence the latitudinal flare distribution.

Despite the potential positional alignment of certain flares with starspot regions, our analysis reveals no clear correlation between flare energy and local spot properties. As shown in Fig.~\ref{fig:fsvsflare}, flare energy does not scale with the spot filling factor computed for the visible hemisphere at the time of each flare. The detected flare energies vary by more than an order of magnitude, even among events associated with similar longitudes or hemispheric spot coverage. This finding is consistent with solar observations, where flare energy exhibits large scatter even within active regions of comparable size \citep{Kazachenko2017}.

Given the limited number of flare events (N = 12), we caution against over-interpreting the apparent lack of correlation. The spot filling factor might constrain the upper envelope of possible flare energies but does not determine the energy of individual flares. A statistically meaningful relationship may only emerge from a much larger flare sample, as emphasized in prior studies \citep{Notsu2019,Namekata2024}. Therefore, continued long-term, high-cadence monitoring of PW And and similar targets is essential to evaluate the interplay between spot coverage, magnetic complexity, and flare energetics.

As discussed in Sect.~\ref{ssec:flares}, the DI+LCI map of PW And reveals a surface configuration characterized by multiple moderately sized spots distributed over a broad range of longitudes. Unlike models based solely on light curve fitting, which often assume a small number of dominant spots, the DI+LCI approach provides a more spatially resolved and realistic depiction of the stellar surface. This configuration is reminiscent of the surface maps of Kepler-17 obtained via the transit method \citet{Namekata2020}, where a similarly dispersed spot distribution was found.

Such a spot configuration helps to explain the relatively smooth shape of the phase-folded light curve and the lack of strong flare phase clustering. In particular, the combined effects of multiple active longitudes and predominantly high-latitude spots, coupled with the star’s moderate inclination, result in persistent visibility of active regions across most rotational phases. Consequently, the occurrence of flares at nearly any phase weakens the expected phase dependence of flare frequency, consistent with previous observations for other young active stars  \citep{Hawley2014,Doyle2018,Ikuta2023}.

Although the detected flare events lie outside the exact time span of the light curve that overlaps with our spectroscopic data (Fig.~\ref{fig:mapvsflare}, middle panel), the relatively stable amplitude of the light curve throughout this period supports the assumption that our DI+LCI map remains a broadly valid representation of the large-scale spot distribution during the flare events.

\section{Summary and conclusions}\label{sec:conclude}
We performed the first simultaneous surface reconstruction of the pre-main-sequence flare star PW And using Doppler imaging and light-curve inversion techniques, combining high-resolution time-series GAOES-RV spectroscopic data from the 3.8 m Seimei Telescope with high-precision photometric observations from TESS. We conclude that the surface spot distribution on PW And is mainly concentrated at mid- to high-latitudes (+30$^{\circ}$ to +90$^{\circ}$), with additional spot structures located at low latitudes on both sides of the equator. Overall, the spotted regions cover approximately 9.9\% of the visible stellar surface. In comparison with previous studies, our reconstructed surface brightness distribution revealed the presence of spots near the equator, which were recovered as a direct result of employing the simultaneous DI+LC inversion method. This approach also enabled a more realistic estimation of the overall spot coverage on the visible stellar surface, compared to 5.4\% when using DI alone.

We studied the temporal changes of the surface distribution of spots, by comparing two DI+LCI reconstructions that are 5.5 rotations (10 days) apart. In common practice, such spectra are often merged to produce a single Doppler image. Motivated by significant changes in light-curve morphology, however, we applied a separate solution for the later dataset despite low spectral phase coverage. The results showed remarkable changes in the spot distribution, which we interpreted as emergence and decay of spots during the 5.5 rotations in between the two epochs.

The detailed simulations performed in this study showed that achieving a reliable and detailed reconstruction of the stellar surface requires the integration of both spectroscopic and photometric datasets \citep[see also][]{Waite2011,Finociety2021,Finociety2023}. Each method alone provides only a partial view, but their complementary combination offers a more accurate view of the surface structure. Photometric light curves are most sensitive to active regions that are positioned so as to produce rotational modulation, in this case near the equator and at low latitudes. However, their sensitivity diminishes at latitudes poleward/northward from the line of sight, since such high-latitude features remain constantly visible and induce only minimal brightness variations due to projection effects set by the axial inclination.

By comparing the reconstructed surface maps with the rotational phases of flares detected by TESS with a temporal shift of one to a few rotations, we investigated the spatial relationship between flares and starspots. The DI+LCI map shows that the active regions that are potentially associated with the rotational phase where flares were detected have two key characteristics: they are primarily concentrated in a mid-to-high latitude band (+30$^{\circ}$ to +60$^{\circ}$), and are simultaneously distributed across multiple longitudes, characterized by several comparable-sized spots. The overall stable photometric amplitude and the observed distribution of flare frequency across a wide range of rotational phases can both be explained by this longitudinally dispersed configuration, which ensures that active regions remain almost continuously visible, although a statistical test of the latter is inconclusive due to the limited sample. As such, the results suggest that the DI+LCI approach is a key step toward more reliably associating flares with their specific surface origins, offering the distinct advantage of resolving spot latitudes and configuration. 

As future studies, we will (1) apply the same method to a young solar analogue and (2) conduct Doppler imaging simultaneously with the H$\alpha$ and Ca II H\&K spectra through the spectroscopic monitoring of flaring active stars with the GAOES-RV \citep{Sato2024} and MId Dispersion Spectrograph for Stellar Activity Research (MIDSSAR), to explore the relation between starspots, chromospheric activity, superflares, and plasma eruptions \citep[e.g.,][]{Namekata2024}.

\begin{acknowledgements}
We thank the referee, Pascal Petit, for his insightful and constructive comments, which led to a substantial improvement of the manuscript. The optical spectroscopic data were obtained through the program number 24B-K-0021 (PI: D.N.) with the 3.8 m Seimei telescope at Okayama Observatory, Kyoto University.
The GAOES-RV project started as a collaboration between Gunma
Astronomical Observatory and Institute of Science Tokyo. Based on the
contract signed between the two parties, GAOES-RV is lent to Institute
of Science Tokyo and operated at the Seimei Telescope.
This study is also based on publically available data obtained by the TESS mission through the MAST data archive at the Space Telescope Science Institute (STScI). Funding for the TESS mission is provided by NASA Science Mission Directorate.
This work is supported by JSPS KAKENHI Grant Numbers JP24H00248 (S.L., K.I., K.N., S.H., and D.N.), JP24K00680 (K.N., S.H., and D.N.), JP24K17082 (K.I.), and JP25K01041 (K.N.).    
\end{acknowledgements}

\bibliographystyle{aa}
\bibliography{biblio}

\begin{appendix}

\section{Simulation using the inferred spot properties of PW And}

\begin{figure*}
 \begin{center}
  \includegraphics[width=15cm]{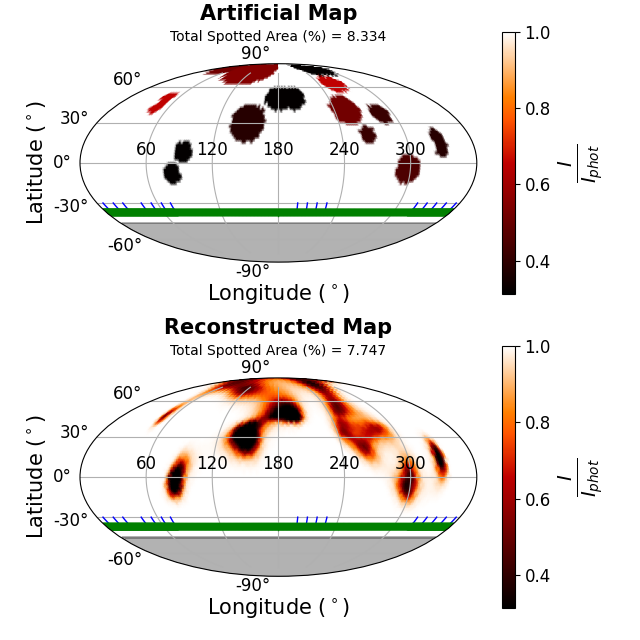}
 \end{center}
 \caption{DI+LCI reconstructed spot map, based on an input map having features similar to the observational reconstruction with the same S/N and spectral phase coverage as in Fig.~\ref{fig:mollmap_lsd} top panel}.\label{fig:dilc_realistic}
\end{figure*}

\begin{table}[ht]
\caption{Properties of the artificial starspots in Fig.~\ref{fig:dilc_realistic}.}
\label{tab:spots}
\centering
\begin{tabular}{ccccc}
\hline\hline
Spot No. & Latitude [$^\circ$] & Longitude [$^\circ$] & Radius [$^\circ$] & Spot Filling Factor \\
\hline
        1  & 10  & 330 & 6  & 0.90 \\
        2  & 20  & 329 & 6  & 0.90 \\
        3  & -5  & 298 & 11 & 0.80 \\
        4  & 40  & 252 & 12 & 0.70 \\
        5  & 21  & 265 & 7  & 0.85 \\
        6  & 37  & 286 & 8  & 0.85 \\
        7  & 63  & 262 & 8  & 0.50 \\
        8  & -8  & 83  & 8  & 0.99 \\
        9  & 8   & 93  & 8  & 0.99 \\
        10 & 42  & 47  & 5  & 0.50 \\
        11 & 51  & 47  & 5  & 0.50 \\
        12 & 80  & 50  & 7  & 0.70 \\
        13 & 79  & 275 & 8  & 0.99 \\
        14 & 75  & 120 & 13 & 0.65 \\
        15 & 50  & 180 & 10 & 0.99 \\
        16 & 50  & 196 & 10 & 0.99 \\
        17 & 30  & 150 & 15 & 0.90 \\
\hline
\end{tabular}
\end{table}

\end{appendix}

\end{document}